\documentclass[a4paper,fleqn,usenatbib,useAMS]{mnras}

\usepackage{mathptmx}

\usepackage[T1]{fontenc}
\usepackage{ae,aecompl}

\usepackage{graphicx}	
\usepackage{amsmath}	
\usepackage{amssymb}	

\newcommand{\logg}{\log\,g}
\newcommand{\teff}{T_{\rm eff}}

\title[The chemical composition of AK\,Sco]{
The chemical composition of the Herbig~Ae SB2 system AK\,Sco (HD\,152404)
}

\stepcounter{footnote}

\author[F. Castelli et al.]{
F.~Castelli$^{1}$\thanks{Deceased.},
S.~Hubrig$^{2}$\thanks{Corresponding author: shubrig@aip.de},
S.~P.~J\"arvinen$^{2}$,
M.~Sch\"oller$^{3}$
\\
$^{1}$Istituto Nazionale di Astrofisica -- Osservatorio Astronomico di Trieste, Via Tiepolo~11, 34131~Trieste, Italy\\
$^{2}$Leibniz-Institut f\"ur Astrophysik Potsdam (AIP), An der Sternwarte~16, 14482~Potsdam, Germany\\
$^{3}$European Southern Observatory, Karl-Schwarzschild-Str.~2, 85748 Garching, Germany
}

\date{Accepted XXX. Received YYY; in original form ZZZ}

\pubyear{2019}

\begin{document}
\label{firstpage}
\pagerange{\pageref{firstpage}--\pageref{lastpage}}
\maketitle

\begin{abstract}
We investigate the stellar atmospheres of the two components
of the Herbig Ae SB2 system AK\,Sco to determine the elements present
in the stars and their abundance.
Equal stellar parameters $\teff=6500$\,K and $\logg=4.5$ were used for both stars.
We studied HARPSpol spectra (resolution 110\,000) that were previously used to state
the presence of a weak magnetic field in the secondary.
A composite synthetic spectrum was compared in the whole observed region
$\lambda\lambda$ 3900--6912\,\AA{} with the observed spectrum.
The abundances were derived mostly from unblended profiles,
in spite of their sparsity,
owing to the complexity of the system
and  to the not negligible $v\,\sin\,i$ of 18\,km\,s$^{-1}$
and 21\,km\,s$^{-1}$ adopted for the two components, respectively.
The identified elements are those typical of stars with  spectral type F\,5\,IV-V,
except for \ion{Li}{i} at 6707\,\AA{} and \ion{He}{i} at 5875.61\,\AA{},
whose presence is related with the Herbig nature of the two stars.
Furthermore, overabundances were determined in both stars for Y, Ba, and La.
Zirconium is overabundant only in the primary,
while sulfur is overabundant outside the adopted error limits only in the secondary.
In contrast to previous results showing a high occurrence rate
of $\lambda$\,Boo peculiarities or normal chemical composition among the Herbig Ae/Be stars,
the abundance pattern of AK\,Sco is similar to that of only few other Herbig stars 
displaying weak Ap/Bp peculiarities.
A few accretion diagnostic lines are discussed.
\end{abstract}




\section{Introduction}
\label{sect:intro}

The combination of knowledge about the magnetic field structure
and the determination of the stellar chemical composition is
very likely the key to constrain theories on star formation
and magnetospheric accretion in intermediate-mass Herbig Ae/Be stars.
These stars are surrounded by active accretion disks and most of the excess
emission present at various wavelengths can probably be ascribed to
the interaction of the disk with a magnetically active star (e.g.\
\citealt{Muzerolle2004}).

Magnetic Herbig~Ae/Be stars are frequently considered as precursors of the magnetic Ap
stars (e.g.\ \citealt{StepienLandstreet2002}; \citealt{Catala2003}). However,
in contrast to chemical peculiar magnetic Ap stars, for which the abundance anomalies are
believed to be produced by mechanisms intrinsic to the stars
themselves, such as radiatively driven diffusion (e.g.\ \citealt{Michaud1970}),
the photospheric material in Herbig~Ae/Be stars
is mixed with circumstellar material originating in a protoplanetary
disk. Therefore, studies of photospheric composition in these
stars can also provide information on protoplanetary material.

Only very few abundance analyses of Herbig Ae/Be stars were carried out in the past.
\citet{Cowley2010}
reported on a chemical composition similar to that in $\lambda$\,Boo stars
in the magnetic Herbig Ae star HD\,101412. Later-on, \citet{Folsom2012}
reported on the abundance
analysis of 20 Herbig Ae/Be stars, concluding that half the stars in their sample
show $\lambda$~Boo chemical peculiarities. Only one star in their study showed weak
Ap/Bp peculiarities and all the remaining stars were chemically normal. In contrast, the
abundance study of the double-lined spectroscopic binary (SB2) system HD\,104237
by \citet{Cowley2013} showed slight enhancements of a
few elements, where the case for Zr was the strongest.
The Herbig~Ae star PDS\,2 was studied by \citet{Cowley2014} and found to have
$\lambda$\,Boo characteristics.

AK\,Sco  (HD\,152404) is a double-lined spectroscopic binary
system formed by two nearly identical Herbig~Ae stars
(F5\,V spectral type) surrounded by a circumbinary disk.
Furthermore, the gap between the stars and the disk is filled by
some gas.
The Herbig nature of AK\,Sco   was first stated by
\citet{HerbigKameswara1972}, who observed 
H$\alpha$ emission with a deep reversal
and \ion{Li}{i} absorption at 6707\,\AA{}
in  Coud\'e spectrograms.
However, according to \citet{Andersen1989}, a follow-up observation with the ESO 1.5\,m
telescope on La~Silla in 1986 did not reveal the presence of
emission lines in the blue spectral region. 
On the other hand, this  observation showed doubling of all photospheric spectroscopic
lines, indicating that AK\,Sco is a spectroscopic binary with approximately equal components.
Subsequent radial-velocity observations
discovered that it is indeed a SB2 system
with a short period (13.6\,days) and large eccentricity ($e=0.47$) \citep{Andersen1989}.

\citet{Alencar2003} reconsidered the orbital parameters and the
physical elements of the system finding a substantial agreement with the
results of \citet{Andersen1989}. Furthermore, \citet{Alencar2003} inferred
the pre-main-sequence (PMS) nature of the secondary from the presence
of the strong \ion{Li}{i} line at 6707\,\AA{} in the spectrum.
A characteristic of AK\,Sco is the light variability, which was shown
to be related to a variable circumstellar obscuration 
\citep{Andersen1989,Alencar2003}
rather than to the orbital motion of the two components.
Recent spectropolarimetric observations
using HARPSpol spectra, performed at six different orbital phases
($\phi=0.946$, 0.950, 0.018, 0.025, 0.090, and 0.169), indicated the presence
of a weak magnetic field of the order of 80\,G  in the
secondary component at the phase 0.090 \citep{Jarvinen2018}.
This discovery raised the question whether chemical
peculiarities like those observed in the main sequence Ap stars can be observed
also in AK\,Sco. 

There are only few studies on the chemical composition of magnetic Herbig~Ae/Be stars.
\citet{Folsom2012} analyzed the abundances in a sample of
21 Herbig~Ae/Be stars. Three of them are affected by the presence of a magnetic field.
They found that the magnetic
Herbig stars do not exhibit a chemical composition remarkably different
from the normal stars. One of them (HD\,101412) displays
$\lambda$\,Boo chemical peculiarities, another (V\,380\,Ori\,A) shows
weak Ap/Bp peculiarities, and  the third one (HD\,190073) was found to be normal.  
The chemical composition of HD\,101412 and HD\,190073 was also studied by \citet{Cowley2010}
and \citet{CowleyHubrig2012}, respectively.
Their results roughly agree with those from \citet{Folsom2012}.
In addition, \citet{Cowley2013} and \citet{Cowley2014}
analyzed the spectra of two other magnetic Herbig~Ae stars, HD\,104237 and
PDS2. The presence of a magnetic field was  dubious for PDS2,
until \citet{Hubrig2015} definitively detected it.

Abundances for fourteen  magnetic (including HD\,104237) and non-magnetic Herbig~Ae/Be stars 
were also presented by \citet{AckeWaelkens2004}. Both \citet{Cowley2013}
and \citet{AckeWaelkens2004}
demonstrated that the abundances of HD\,104237 are slightly peculiar with
enhanced elements Y, Zr, Ba, and La.
The agreement between the results from the two studies concerning these elements
is within 0.07\,dex. In contrast, the chemical composition of PDS2 was found to be similar to
that of the $\lambda$\,Boo stars.

In this paper we use HARPSpol spectra
covering the 3900--6912\,\AA{} region to investigate the composite spectrum of AK\,Sco
with the aim to increase the number of abundance analyses
of magnetic Herbig~Ae/Be stars.

\section{Observations}
\label{sect:obs}

\begin{figure}
\centering
\includegraphics[width=0.45\textwidth]{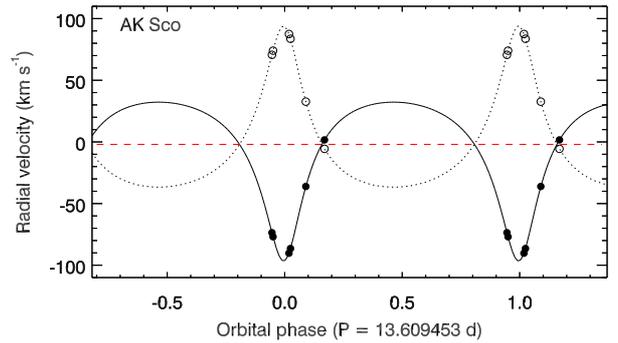}
\caption{Radial velocity curve  from the orbital solution from
\citet{Alencar2003}.
The circles indicate the phases of our observations.
Star A (black filled circles) has its minimum velocity at phase 0.00.} 
\label{fig:f1}
\end{figure}

We used the same HARPSpol spectropolarimetric observations of AK\,Sco
and the same reduced spectra that are described in \citet{Jarvinen2018}.
Here we recall that the observations cover six orbital phases (0.946, 0.950,
 0.018, 0.025, 0.090, 0.169) of the binary
 system. The covered orbital phases (calculated using the orbital solution
 from \citet{Alencar2003} are presented in Fig.~\ref{fig:f1}.
 While the observed wavelength interval is 3780$-$6912\,\AA{}, with a small gap
between 5259 and 5337\,\AA{}, the region useful for the analysis starts at 3900\,\AA{} because
the noise is too strong at
lower wavelengths. The resolving power is about 110\,000.

In this work, the normalization to the continuum performed
by \citet{Jarvinen2018} was adjusted at intervals of 
6\,\AA{} by comparing observed and computed spectra. The continuum level
was lowered from about 30\% at 4000\,\AA{}, to about 5\% at 4600\,\AA{}, to
0.05$-$0.25\% for $\lambda$$>$ 5000\,\AA{}. 
In all the spectra, except for that at the phase 0.169, the spectral lines
of both components are well separated.
The variability was studied by using all six spectra observed at
the different epochs. 

\section{Stellar parameters and synthetic spectra} 
\label{sect:parameters}

\begin{table}
  \caption[ ]{Observed phases, difference in radial velocity between
    the two components, the radial velocity of the primary, and
    the luminosity fraction (in \%) observed  from the secondary.} 
\label{tab:t1}
\centering
\begin{tabular}{crlccc}
\hline\noalign{\smallskip}
\multicolumn{1}{c}{}&
\multicolumn{1}{c}{Date}&
\multicolumn{1}{c}{Phase}&
\multicolumn{1}{c}{$v_{\rm r}$(B)-$v_{\rm r}$(A)}&
\multicolumn{1}{c}{$v_{\rm r}$(A)}&
\multicolumn{1}{c}{$\frac{L({\rm B}_{\rm obs})}{L({\rm B}_{\rm tot})}$}
\\
\noalign{\smallskip}
\multicolumn{1}{c}{}&
\multicolumn{1}{c}{}&
\multicolumn{1}{c}{}&
\multicolumn{1}{c}{[km\,s$^{-1}$]}&
\multicolumn{1}{c}{[km\,s$^{-1}$]}&
\multicolumn{1}{c}{[\%]}
\\
\hline\noalign{\smallskip}
  1&  2016-06-15 & 0.946 & 138 &$-$70 & 10\\
  2&  2017-06-04 & 0.950 & 146 &$-$72.5 & 50\\
  3&  2016-06-18 & 0.018 & 180 &$-$90   & 50\\
  4&  2017-06-05 & 0.025 & 177 &$-$87   & 80\\ 
  5&  2017-06-06 & 0.090 & 74  &$-$36   &100\\
  6&  2017-06-07 & 0.169 & 1.  & 0.   &100\\
\hline
\noalign{\smallskip}
\end{tabular}
\end{table}

The abundances were derived from the spectrum observed at the phase $\phi=0.090$, i.e.
from the spectrum for which the presence of a weak magnetic field was established
\citep{Jarvinen2018}. Furthermore, at this phase the two stars are well separated
with a radial velocity  shift $\Delta v=74$\,km\,s$^{-1}$ (see Table~\ref{tab:t1}).

The double-line spectrum, the similar intensity of the lines from the two stars,
and the rather high rotational velocities
(18\,km\,s$^{-1}$ and 21\,km\,s$^{-1}$, respectively)
increase the number of  blended profiles, so that it is very difficult
to pick up lines that do not have any contamination  with
lines either from the same star or  from the companion.
Therefore, the whole
analysis was performed with the synthetic spectrum method.
Unblended lines are rare, so that the abundance for a given
element was determined from the profiles of few single lines, when available.
Then the abundance was checked on blended lines having the element as main or
sometimes even minor component of the blend.

The atomic and molecular line lists adopted  to compute the synthetic spectra 
were  produced by Castelli\footnote{http://wwwuser.oats.inaf.it/castelli/linelists.html},
who assembled the line lists from the GFNEW directory \citep{Kurucz2018}, available at
the Kurucz website\footnote{http://kurucz.harvard.edu/linelists.html}, with line lists
taken from the literature for some ions missing in the Kurucz data. 
Furthermore, if needed, the $\log gf$ values in the Kurucz line lists
were replaced by values either extracted from various literature sources
or obtained by fitting the solar synthetic spectrum to the
observed solar flux Atlas from \citet{Kurucz2005a}.
For numerous lines, Van der Waals broadening was computed according to
the \citet{AnsteeOMara1995} theory. The $\gamma_{\rm VdW}$ and $\alpha$ parameters
were taken from the \citet{Barklem2000} tables.

We adopted for both stars $\teff=6500$\,K from \citet{Andersen1989},
who inferred this temperature on the basis of the F5\,V spectral type.
They used the $\teff$ $-$ spectral type calibration given by \citet{Bessell1979} and
by \citet{Popper1980}. For this value of $\teff$, \citet{Alencar2003} 
determined the system parameters assuming nearly identical components,
in particular
$M=1.35\pm0.07$\,M$_{\odot}$, $R=1.59\pm0.35$\,R$_{\odot}$, and
$v\,\sin\,i=18.5\pm1$\,km\,s$^{-1}$, provided that the stellar inclination is
between 65 and 70 degrees.
The gravity, deduced from the mass and radius given above,
is $\logg=4.16\pm0.25$.

\begin{figure}
\centering
\resizebox{0.45\textwidth}{!}{\rotatebox{90}{\includegraphics[56,228][514,790]{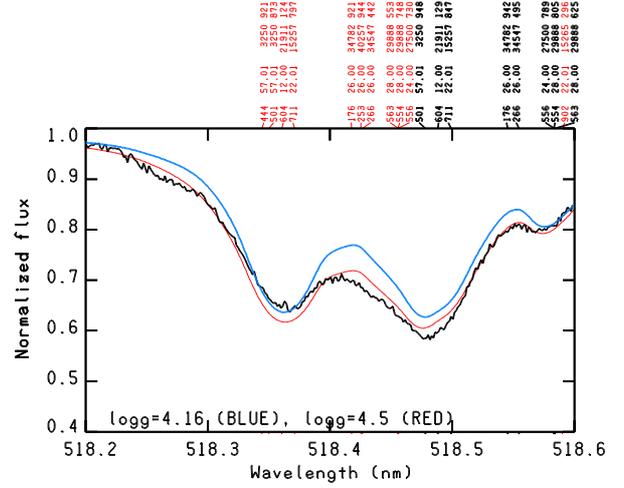}}}
\caption{Comparison of the \ion{Mg}{i} profile at 5183.604\,\AA{}
  observed in the composite spectrum   at the phase 0.090 (black line)
  with  two spectra computed with  solar magnesium abundance, the same $\teff=6500$\,K, and
  two different  $\logg = 4.16$ (blue line)  and 4.5 (red line).
  The red and black  labels identify the
  lines observed in the primary and in the secondary, respectively.  
  The wavelength scale corresponds to the laboratory wavelengths  of the
  primary.  The identification labels of the secondary have to be read on
  wavelength scale corresponding to the laboratory wavelengths of the
  secondary (i.e.\ 5184.604 for the secondary has to be read 5183.604). 
}
\label{fig:f2}
\end{figure}

In order to check the gravity obtained from  mass
and radius spectroscopically, we used the wings of the \ion{Mg}{i} triplet at
5167.321, 5172.684, and especially 5183.604\,\AA{},
because it does not show the slight redshift
as the other two lines do.
At first, we determined  the Mg abundance $\log(N_{\rm Mg}/N_{\rm tot})=-4.55$
for $\teff=6500$\,K and $\logg=4.16$
using the \ion{Mg}{ii} line at 4481\,\AA{}, which is almost independent of gravity.
For this abundance and  $\logg=4.16$
the computed wings of the \ion{Mg}{i} triplet were too narrow.
To improve the agreement between the observed and computed
\ion{Mg}{i} profiles we increased the gravity to $\logg=4.5$ (Fig.~\ref{fig:f2}).
We note that when the line data were checked on the solar spectrum,
we decreased for the \ion{Mg}{i} triplet the parameter $\log\gamma_{\rm VdW}$
from $-7.27$, as given by \citet{Barklem2000}, to $-7.37$.  
Probably, a different solar model than that used by us would have given  good
agreement   between observed  and computed wings of the solar \ion{Mg}{i} triplet
for $\log\gamma_{\rm VdW}=-7.27$.
For the Sun, we adopted an ATLAS9 model with parameters $\teff=5777$\,K and $\logg=4.4377$,
and turbulent velocity $\xi=1$\,km\,s$^{-1}$. This model was used because
 it is consistent with the ATLAS9 model adopted to analyze AK\,Sco.

An ATLAS9 model atmosphere with parameters $\teff=6500$\,K and $\logg=4.5$
from the updated
\citet{CastelliKurucz2004}\footnote{http://wwwuser.oats.inaf.it/castelli/grids.html} grid was used to
compute synthetic spectra for both components by means of the SYNTHE code \citep{Kurucz2005b}.
On the basis of the comparison of
the observed and computed profiles, the microturbulent velocity
was estimated to be $\xi=1$\,km\,s$^{-1}$ and $\xi=2$\,km\,s$^{-1}$ for the
primary and the secondary, respectively. The rotational velocity
$v\,\sin\,i$ was determined from the comparison of observed and computed
profiles of numerous stellar features (e.g.\ \ion{Mg}{ii} 4481\,\AA{}).
We adopted the values 18\,km\,s$^{-1}$ and 21\,km\,s$^{-1}$ for the
primary and the secondary, respectively.  Finally, the computed profiles were broadened 
for a Gaussian instrumental profile, corresponding to the resolving power
110\,000  of the HARPSpol spectrum.

The composite spectrum was obtained with the BINARY code \citep{Kurucz1995}.
As described by \citet{Cowley2013}, the  spectra of the two components are
shifted in respect to each other in accordance with
the observed radial velocity difference. 
Then, they were weighted with the luminosity ratio and
added together. The adopted luminosity ratio is 1.0, 
corresponding to  a radii ratio  equal to 1.0  \citep{GomezdeCastro2009}.
Because not only all the accretion diagnostic lines, but also  photospheric lines
show intensity variations over the observing nights \citep{Jarvinen2018},
we investigated whether the
photospheric line variability may be caused by abundance changes.
By comparing observed and computed spectra at different phases, we concluded that
the variability of the photospheric lines should be mostly related to
the presence of the circumbinary disk, which obscures the secondary
component with dust clouds with different densities at different phases.
We assumed ``a priori''  that there is no obscuration at the phase 0.090,
so that we adopted as stellar abundances those derived at this phase.
For the other phases, we determined for the secondary the fraction
in percent of
the observed luminosity to its total luminosity 
on the basis of the \ion{Li}{i} profile at 6707\,\AA{}. We assumed
that the abundance of \ion{Li}{i} in all phases is the same as that at
the phase 0.090. Table~\ref{tab:t1} summarizes for the different phases
the radial velocity difference for the two stars,
the radial velocity of the primary as determined from the spectra, and
the observed luminosity fraction (in \%)  of the secondary.

\section{Identification and abundances}
\label{sect:abundances}

\begin{table*}
 \caption{Abundances $\log$(N$_{elem}$/N$_{tot}$) of the identified elements
 in AK\,Sco. 
 Solar abundances are from
\citet{Asplund2009},
\citet{Scott2015a,Scott2015b},
and \citet{Grevesse2015}.
The values in round brackets in Cols.~2 and 4 denote the numbers of lines
used in the spectral analysis of that ion.
The values in square brackets in Cols.~3 and 5 denote the differences
between the abundance found in AK\,Sco and the solar abundance.
}
\label{tab:t2}
\centering
\begin{tabular}{llclccccccccccccccc}
\hline\noalign{\smallskip}
\multicolumn{1}{c}{Element}&
\multicolumn{2}{c}{A component} &
\multicolumn{2}{c}{B component} &
\multicolumn{1}{c}{Sun}
 \\
 &
\multicolumn{2}{c}{($\teff=6500$\,K, $\log g=4.5$)} &
\multicolumn{2}{c}{($\teff=6500$\,K, $\log g=4.5$)} &
 \\
\hline\noalign{\smallskip}
\ion{Li}{i}    & $-$8.70 (1)           & [+2.29]   & $-$8.50 (1)           & [+2.49]   & $-$10.99 \\
\ion{C}{i}     & $-$3.61 (1)           & [0.00]    & $-$3.61 (1)           & [0.00]    & $-$3.61 \\
\ion{O}{i}     & $-$3.35 (3)           & [0.00]    & $-$3.35 (3)           & [0.00]    & $-$3.35 \\
\ion{Na}{i}    & $-$5.93 (2)           & [$-$0.10] & $-$5.78$\pm$0.05 (3)  & [+0.05]   & $-$5.83 \\
\ion{Mg}{i}    & $-$4.55 (2)           & [$-$0.10] & $-$4.45 (3)           & [+0.00]   & $-$4.45 \\
\ion{Mg}{ii}   & $-$4.55 (1)           & [$-$0.10] & $-$4.45 (1)           & [+0.00]   & $-$4.45 \\
\ion{Al}{i}    & $-$5.61 (2)           & [0.00]    & $-$5.61 (2)           & [+0.00]   & $-$5.61 \\
\ion{Si}{i}    & $-$4.40$\pm$0.07 (12) & [+0.13]   & $-$4.31$\pm$0.06 (7)  & [+0.22]   & $-$4.53 \\
\ion{Si}{ii}   & $-$4.40 (1)           & [+0.13]   & $-$4.40 (1)           & [+0.13]   & $-$4.53 \\
Si$_{\rm tot}$ & $-$4.40$\pm$0.07      & [+0.13]   & $-$4.33$\pm$0.07      & [+0.20]   & $-$4.53 \\
\ion{S}{i}     & $-$4.72$\pm$0.10 (2)  & [+0.20]   & $-$4.62$\pm$0.10 (2)  & [+0.30]   & $-$4.92 \\
\ion{Ca}{i}    & $-$5.70$\pm$0.10 (15) & [+0.02]   & $-$5.84$\pm$0.08 (12) & [$-$0.12] & $-$5.72 \\
\ion{Sc}{ii}   & $-$8.98 (3)           & [$-$0.10] & $-$8.68 (2)           & [+0.20]   & $-$8.88 \\
\ion{Ti}{i}    & $-$7.14$\pm$0.04 (3)  & [$-$0.03] &                       &           & $-$7.11 \\
\ion{Ti}{ii}   & $-$6.88$\pm$0.09 (6)  & [+0.23]   & $-$7.14$\pm$0.09 (3)  & [$-$0.03] & $-$7.11 \\
Ti$_{\rm tot}$ & $-$6.96$\pm$0.15      & [+0.15]   & $-$7.14$\pm$0.09      & [$-$0.03] & $-$7.11 \\
\ion{V}{i}     & $-$8.15 (2)           & [0.00]    &                       &           & $-$8.15 \\
\ion{Cr}{i}    & $-$6.41$\pm$0.12 (4)  & [+0.01]   & $-$6.42 (3)           & [0.00]    & $-$6.42 \\
\ion{Cr}{ii}   & $-$6.22$\pm$0.20 (2)  & [+0.20]   & $-$6.35$\pm$0.04 (3)  & [+0.07]   & $-$6.42 \\
Cr$_{\rm tot}$ & $-$6.34$\pm$0.17      & [+0.08]   & $-$6.39$\pm$0.02      & [+0.03]   & $-$6.42 \\
\ion{Mn}{i}    & $-$6.62 (3)           & [0.00]    & $-$6.62 (2)           & [0.00]    & $-$6.62 \\
\ion{Fe}{i}    & $-$4.66$\pm$0.15 (12) & [$-$0.11] & $-$4.51$\pm$0.17 (11) & [+0.06]   & $-$4.57 \\
\ion{Fe}{ii}   & $-$4.42 (2)           & [+0.15]   & $-$4.44$\pm$0.06 (5)  & [+0.13]   & $-$4.57 \\
Fe$_{\rm tot}$ & $-$4.62$\pm$0.16      & [$-$0.05] & $-$4.49$\pm$0.15      & [+0.08]   & $-$4.57 \\
\ion{Co}{i}    &                       &           & $-$6.91 (1)           & [+0.20]   & $-$7.11 \\
\ion{Ni}{i}    & $-$5.84$\pm$0.08 (9)  & [0.00]    & $-$5.74$\pm$0.17 (8)  & [+0.14]   & $-$5.84 \\
\ion{Cu}{i}    & $-$7.86 (1)           & [0.00]    & $-$7.86 (1)           & [+0.00]   & $-$7.86 \\
\ion{Zn}{i}    & $-$7.38 (2)           & [+0.10]   & $-$7.58 (2)           & [$-$0.10] & $-$7.48 \\
\ion{Sr}{i}    &                       &           & $-$9.00 (1)           & [+0.21]   & $-$9.21 \\
\ion{Sr}{ii}   & $-$9.10 (1)           & [+0.11]   & $-$8.90 (1)           & [+0.29]   & $-$9.21 \\
Sr$_{\rm tot}$ & $-$9.10               & [+0.11]   & $-$8.95$\pm$0.05      & [+0.24]   & $-$9.21 \\
\ion{Y}{ii}    & $-$9.40 (2)           & [+0.40]   & $-$9.50 (1)           & [+0.30]   & $-$9.83 \\
\ion{Zr}{ii}   & $-$9.05 (2)           & [+0.40]   & $-$9.45 (1)           & [0.00]    & $-$9.45 \\
\ion{Ba}{ii}   & $-$9.51 (1)           & [+0.28]   & $-$9.41 (1)           & [+0.38]   & $-$9.79 \\
\ion{La}{ii}   & $-$10.22$\pm$0.15 (2) & [+0.71]   & $-$10.57 (1)          & [+0.36]   & $-$10.93 \\
\ion{Ce}{ii}   & $-$10.46 (1)          & [+0.00]   &                       &           & $-$10.46 \\
\ion{Nd}{ii}   & $-$10.62 (1)          & [+0.00]   & $-$10.62              & [+0.00]   & $-$10.62 \\
\hline\noalign{\smallskip}
\end{tabular}
\end{table*}

The whole available spectrum from 3900\,\AA{} to 6900\,\AA{} was synthetized
and compared with the HARPSpol spectrum.
In order to derive the abundances, we analyzed
the profiles of the lines  listed in the Appendix.
The comparison of the observed spectrum with the spectrum computed
with the final abundances  listed in Table~\ref{tab:t2} is available
from Castelli's webpage\footnote{http://wwwuser.oats.inaf.it/castelli/stars/AK\_Sco/ AK\_Sco.html}.
We identified
in both stars: \ion{H}{i}, \ion{Li}{i}, \ion{C}{i}, \ion{O}{i},
\ion{Na}{i},
\ion{Mg}{i}, \ion{Mg}{ii}, \ion{Al}{i}, \ion{Si}{i}, \ion{Si}{ii},
\ion{S}{i}, \ion{Ca}{i}, \ion{Ca}{ii},
\ion{Sc}{i}, \ion{Sc}{ii}, \ion{Ti}{i}, \ion{Ti}{ii}, \ion{V}{i}, \ion{V}{ii},
\ion{Cr}{i}, \ion{Cr}{ii}, \ion{Mn}{i}, \ion{Fe}{i}, \ion{Fe}{ii}, \ion{Co}{i},
\ion{Ni}{i}, \ion{Cu}{i}, \ion{Zn}{i}, \ion{Sr}{i}, \ion{Sr}{ii}, \ion{Y}{ii},
\ion{Zr}{ii}, \ion{Ba}{i}, \ion{Ba}{ii}, \ion{La}{ii}, \ion{Ce}{ii}, and \ion{Nd}{ii}.
In addition, \ion{He}{i} at 5876.61\,\AA{} is observed according to the Herbig
nature of the studied stars. We note that for a temperature of 6500\,K,
\ion{He}{i} is not predicted,
unless it is unreasonably overabundant.
Predicted marginal contributions from CH and CN are
not in conflict with the observations.

Most lines for all elements are well reproduced in the observed spectra 
by solar abundances for both components, except for \ion{Li},  \ion{Sr}, \ion{Y}, \ion{Ba},
and \ion{La}. Zirconium is overabundant only in the primary. Sulfur is 
overabundant by 0.3\,dex in the secondary, while we considered its overabundance of 0.2\,dex
in the primary as solar abundance within the error limits (see Table~\ref{tab:t2}).
The adopted solar abundance for neodymium is a lower limit, because we can not exclude
some overabundance also for this element. While some lines are well fitted
by solar abundances, some others are computed too weak.
These discordant results  can be related to the weakness and the blending of the
\ion{Nd}{ii} lines as well as with the adopted $\log\,gf$ values.
The solar abundances listed in Col.\,6 of Table~\ref{tab:t2} are from \citet{Asplund2009}, 
   \citet{Scott2015a},   \citet{Scott2015b}, and 
   \citet{Grevesse2015}.

Some lines are affected by a variable additional
absorption that has becomes recognizable due to a slightly redshifted
wavelength or a too
broad profile as compared with the computed one. In some cases, the observed core
is flatter than that  predicted by the synthetic spectrum.
Abundances from these lines may differ from those derived from other lines of the same element.

The comparison between the observed spectra at  all phases with the 
computed spectrum indicates that for all elements most lines are well
fitted by the same adopted abundances. Therefore, the variability observed and discussed
in previous works has to be ascribed to the relative position of the two components, to
the circumbinary disk, to the presence of some gas  between the disk and the components, 
rather than to abundance spots over the stellar surfaces.

The main uncertainty sources in the abundance determination are the
location of the continuum level, the $\log\,gf$ values, and missing lines
in the line lists. For weak and medium-strong lines a
change in the continuum level by 10$\%$ results in an uncertainty of
about 0.1\, dex for the abundance. We assumed a total error
of $\pm$0.20\,dex as the maximum error for the adopted abundances.

\section{Line profiles not well reproduced by the LTE synthetic spectrum}
\label{sect:not_reproduced}

While large parts of the spectrum are well reproduced by
the LTE synthetic spectrum using the abundances listed in Table~\ref{tab:t2},
there are some profiles which require more sophisticated models and
methods to be explained in a satisfactory way. They are mostly features
affected by magnetospheric activity, such as the Balmer profiles,
\ion{He}{i} 5875.61\,\AA{}, and the \ion{Na}{i}\,D lines,
which were also discussed by \citet{Pogodin2012}.
In addition, also the \ion{Ca}{ii} K and H lines, 
the strongest lines of \ion{Mg}{i}, and some lines of \ion{Fe}{i}
show signatures of stellar activity.
The list of the not well reproduced line profiles is given below.

{\em Balmer lines:}
 Only H$\alpha$ shows emission. It is present in all six phases.
The profiles differ
each from the other for both the emission and absorption contributions,
and  differ also from the profiles displayed
by \citet{Alencar2003}, which were observed
at phases (i.e.\ $\phi=0.05$, 0.17, 0.57, and 0.81)
different from those we studied in this paper.
The H$\beta$ profiles are
in absorption in all observed phases, although they display
a strong variability.  They 
 are formed by a strong deep absorption core,
not predicted by the model, and by an additional absorption, redshifted by a
few Angstrom, which is
observable in all phases, except for the phase at 0.169, where the
observed and computed central wavelengths coincide. As a consequence,
except for the phase at 0.169, the whole observed H$\beta$ profile is
redshifted and broader than that  predicted by the model.
The H$\beta$ profiles discussed here are similar to those shown by \citet{Alencar2003}.
The same behaviour can be observed also in the
H$\gamma$ and H$\delta$ profiles, although to a lesser extent
than that observed for H$\beta$.

\begin{figure}
\centering
\resizebox{0.45\textwidth}{!}{\rotatebox{90}{\includegraphics[20,372][562,830]{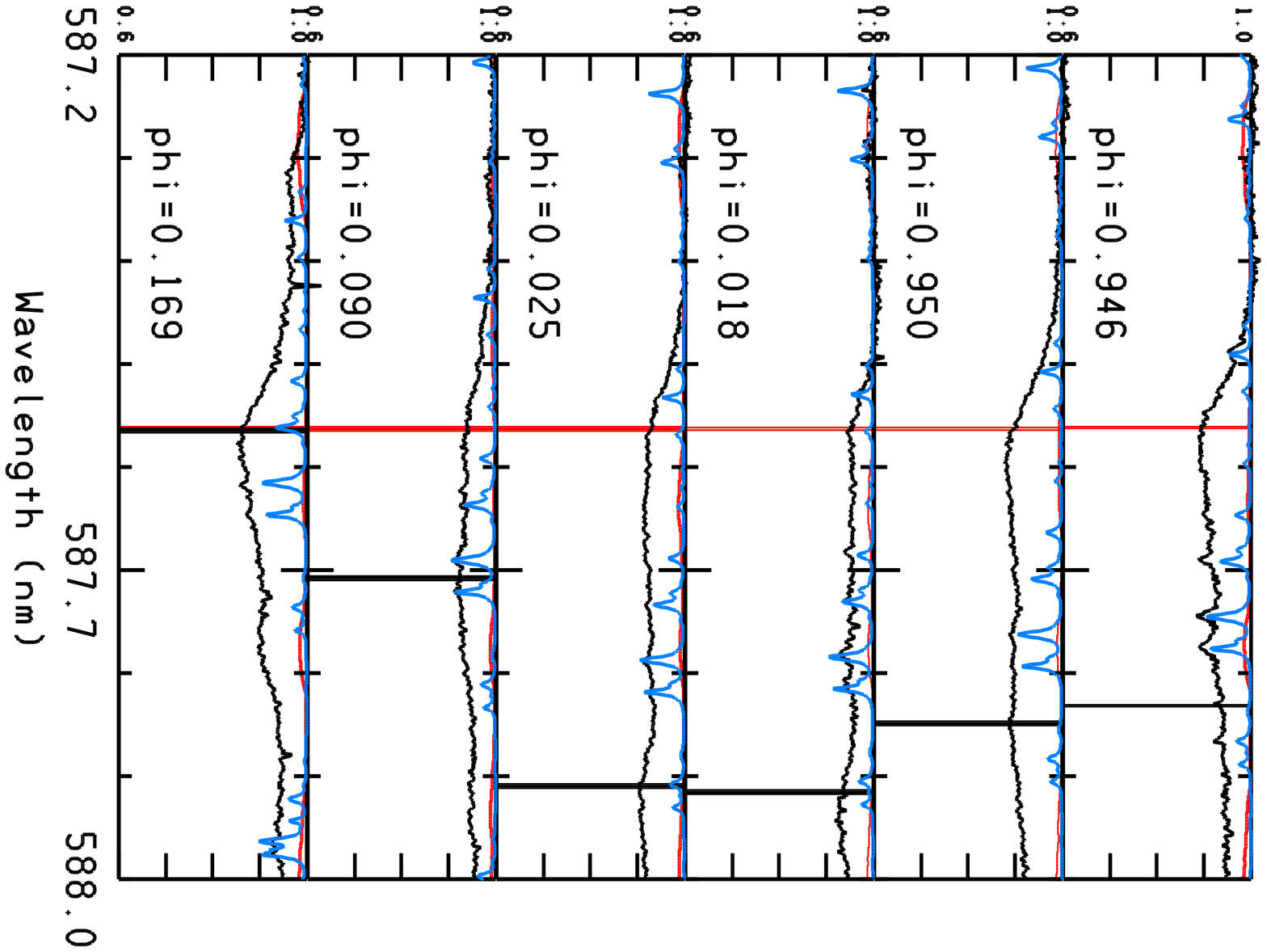}}}
\caption{The \ion{He}{i} line at 5875.6\,\AA{} line arising in the magnetospheric accretion flow
  observed at six different phases in AK\,Sco.
  The red and black vertical lines mark the position of \ion{He}{i} in
  the primary and in the secondary, respectively.
The sharp blue lines indicate contamination by telluric lines.  
 }
\label{fig:f3}
\end{figure}

{\em \ion{He}{i}}: The line of \ion{He}{i} at 5875.61\,\AA{} is 
well observable in both components as a weak variable absorption (Fig.~\ref{fig:f3}). 
It can not be predicted by the synthetic spectrum, whichever helium
abundance is adopted.
No other \ion{He}{i} lines are present in the HARPSpol spectrum. \citet{reiter2018}
studied \ion{He}{i} $\lambda$10830 line profiles in a sample of Herbig Ae/Be stars and reported that
in the near-IR spectrum of AK\,Sco  this line appears partially in emission.

\begin{figure*}
\centering
\resizebox{0.85\textwidth}{!}{\rotatebox{90}{\includegraphics[15,-10][564,828]{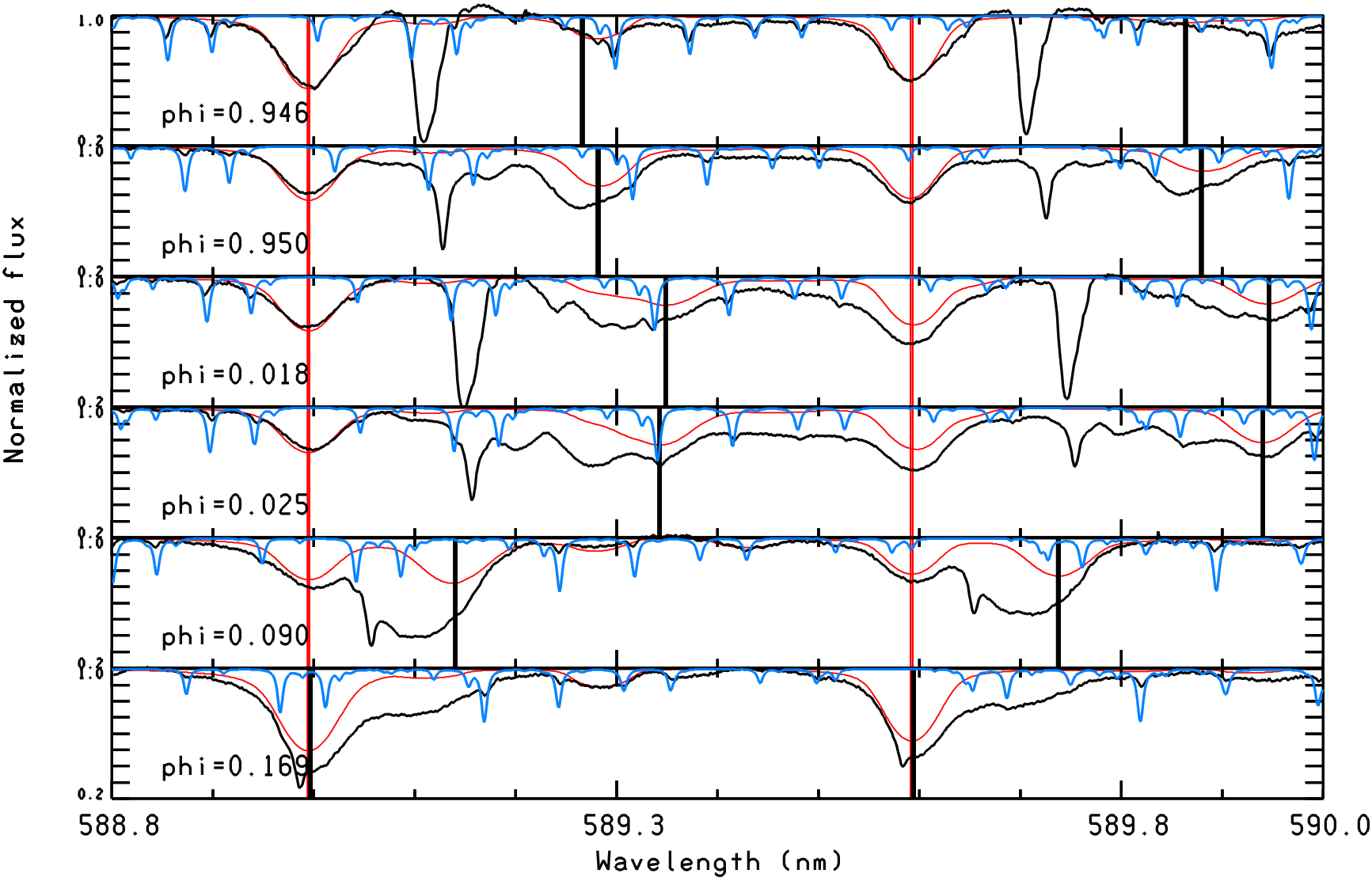}}}
\caption{Comparison of the \ion{Na}{i} doublet at 5889.95\,\AA{} and
  5895.920\,\AA{}, observed at different phases, with the synthetic spectrum.
  The wavelength scale of the primary coincides with the laboratory wavelengths.
  The red and black vertical lines mark the position of the \ion{Na}{i} lines
  in  the primary and in the secondary, respectively.
The sharp blue lines indicate contamination by telluric lines.  
 }
\label{fig:f4}
\end{figure*}

\begin{figure*}
\centering
\resizebox{0.85\textwidth}{!}{\rotatebox{90}{\includegraphics[15,-10][564,828]{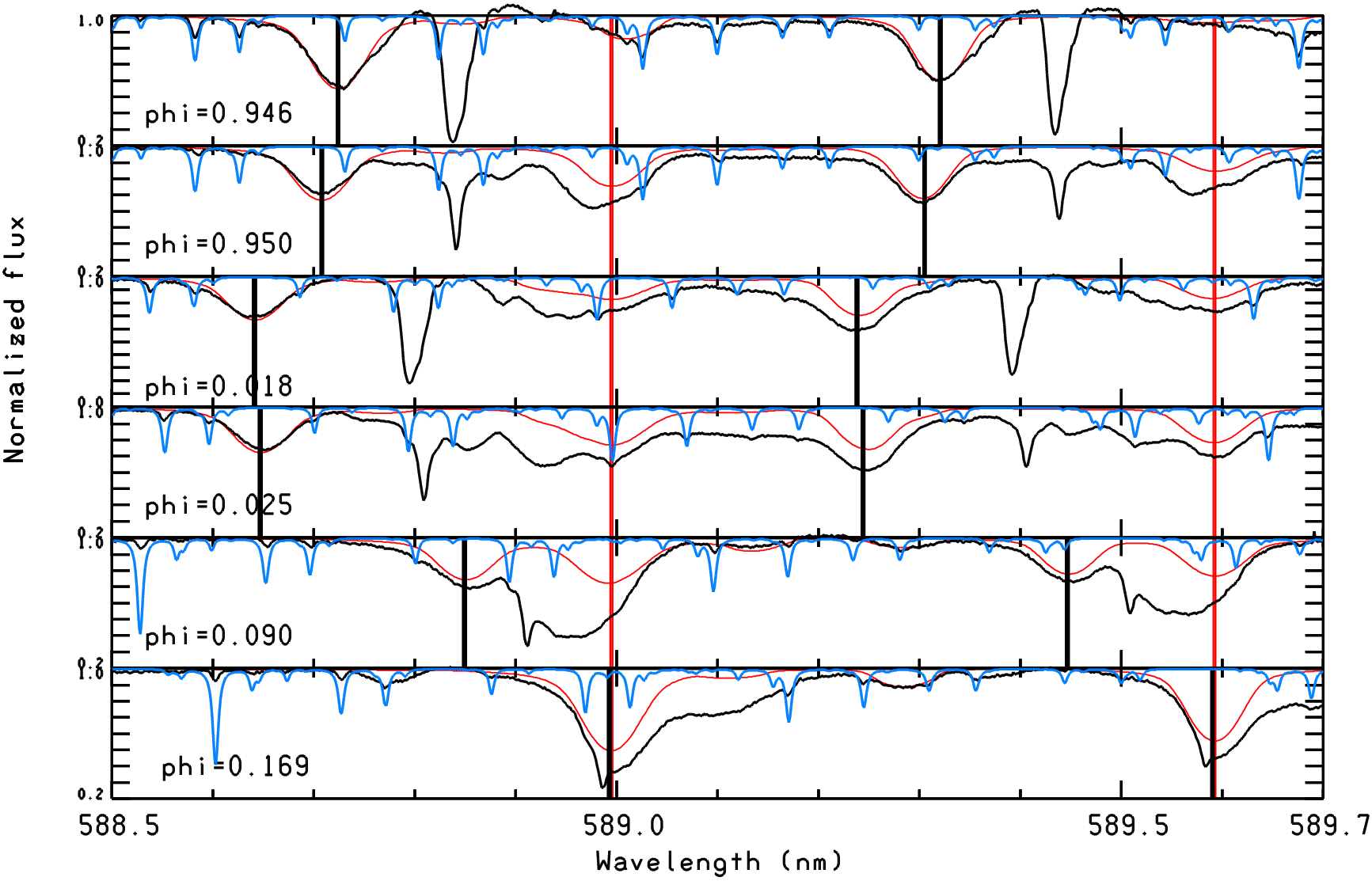}}}
\caption{Comparison of the \ion{Na}{i} doublet at 5889.95\,\AA{} and
  5895.920\,\AA{}, observed at different phases, with the synthetic spectrum.
  The wavelength scale of the secondary coincides with the
  laboratory wavelengths.
  The red and black vertical lines mark the position of the \ion{Na}{i}
  lines in the secondary and in the primary, respectively.
The sharp blue lines indicate contamination by telluric lines.  
 }
\label{fig:f5}
\end{figure*}

{\em \ion{Na}{i}:} The \ion{Na}{i} doublet displays a composite structure.
In the primary (Fig.~\ref{fig:f4}), the line observed at 5889.950\,\AA{} is well predicted  
at  the phases 0.946, 0.950, 0.018, and 0.025, while it is stronger
than the computed one at the phases 0.090 and 0.169. The other line
at 5895.920\,\AA{} is well predicted at the phases 0.946 and 0.950, but it is
computed too weak for all the other phases.
In the secondary (Fig.~\ref{fig:f5}), the \ion{Na}{i} doublet is well
reproduced only at the phase 0.946,
while in all other phases the observed line is stronger than the computed
one, especially at the phases 0.090 and 0.169.
The profiles
are affected by a blue-shifted strong broad component,  probably due to
the magnetospheric interaction with the accretion disk.
Narrow interstellar \ion{Na}{i} lines
are also present in the spectrum in all the observed phases.

{\em \ion{Mg}{i}:} In the primary, the
line core is weaker than the computed one in almost all oberved lines
($\lambda\lambda$ 4702.091, 5167.321, 5172.684, 5183.604, and 5228.405\,\AA{}). 
Furthermore, the lines at 5167.321
and 5172.684\,\AA{} are redshifted by 1.5\,km\,s$^{-1}$, corresponding to
$\Delta\lambda= 0.025$\,\AA{}.
In the secondary, the \ion{Mg}{i} profiles are well reproduced both in shape
and position.

{\em \ion{Al}{i}:} The resonance line at 3944.006\,\AA{} is redshifted by
3\,km\,s$^{-1}$, corresponding to $\Delta\lambda=0.04$\,\AA{}.

{\em \ion{Si}{i}:} The \ion{Si}{i} line at 6237.319\,\AA{} is variable. It
is computed too strong in all phases. 

{\em \ion{Ca}{i}:} Only in the secondary, a few lines of \ion{Ca}{i}
are stronger than the predicted ones. Among them, the line at 4226.728\,\AA{}
is blended, those at 5590.114 and 6122.217\,\AA{} display an observed core stronger
that the computed one, while the line at 6717.081\,\AA{} is either blended
with an unknown component, or is affected by an additional redshifted
absorption. This line is unshifted at the phase 0.090 and redshifted at the other
phases.

\begin{figure*}
\centering
\resizebox{0.85\textwidth}{!}{\rotatebox{90}{\includegraphics[18,34][557,794]{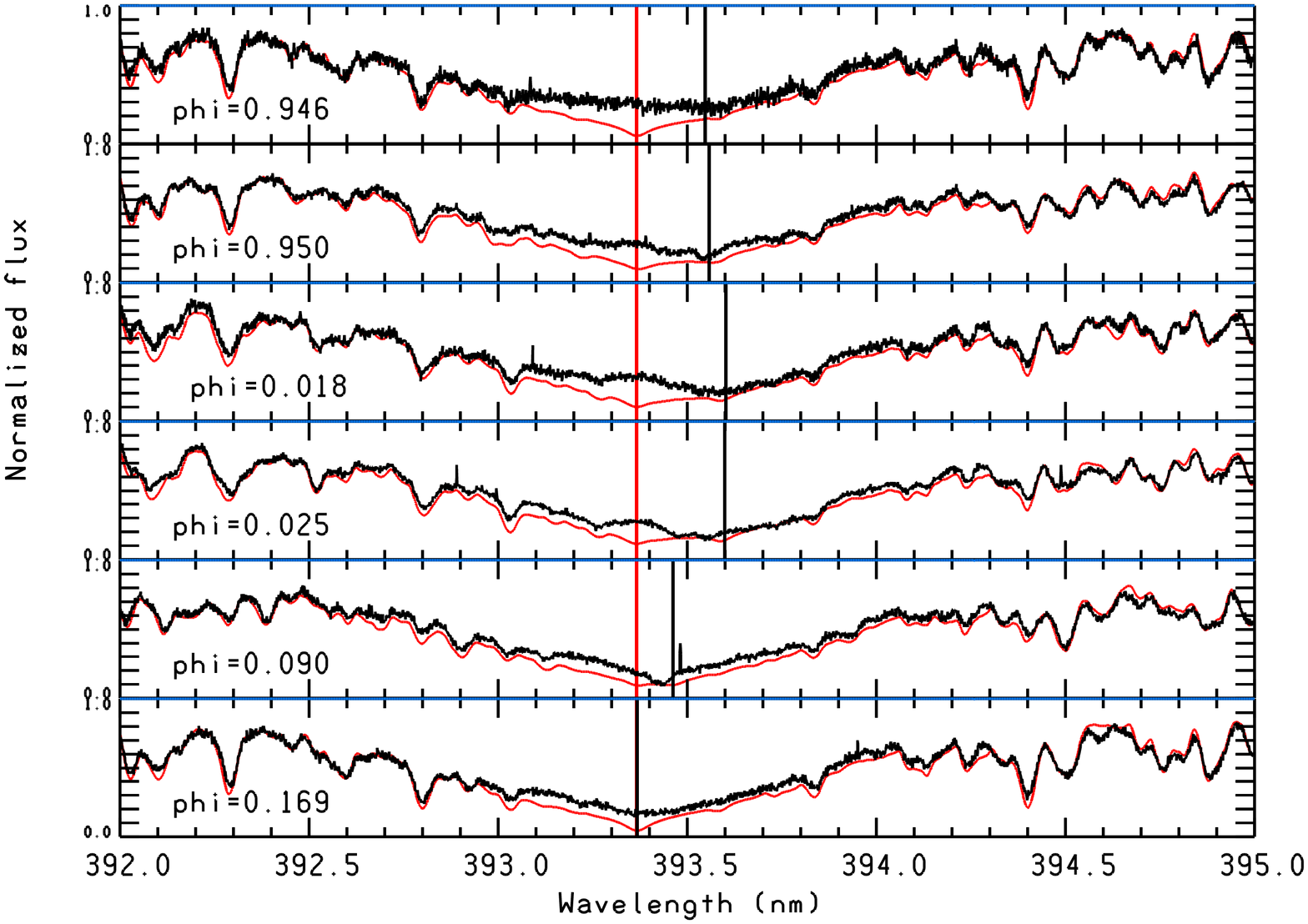}}}
\caption{Comparison with the synthetic spectrum of the \ion{Ca}{ii} K line at 3933.664\,\AA{},
  observed at different phases.
  The wavelength scale of the primary coincides with the
  laboratory wavelengths.
  The red and black vertical lines mark the position of the \ion{Ca}{ii}
  K line in the primary and in the secondary, respectively.
 }
\label{fig:f6}
\end{figure*}

{\em \ion{Ca}{ii}:} For both  \ion{Ca}{ii} K and H profiles the core is flat and weaker than
the computed one with a different shape at different phases
(Fig.~\ref{fig:f6}).

{\em \ion{Fe}{i}:} Several \ion{Fe}{i} lines display an observed core weaker than the computed one.
Examples are the strong lines of \ion{Fe}{i} at 4920.502, 4957.596, 4991.268, 5007.274,
5110.413, 5367.465, 5424.067, and 5445.042\,\AA{}. 
Other \ion{Fe}{i} lines seem to be double, or both double and redshifted.
For instance, for the primary, the observed core of the \ion{Fe}{i}
profiles at 5615.644\,\AA{} is weaker than the
computed one and is redshifted by 0.01\,\AA{} at the phase 0.025.

{\em \ion{Fe}{ii}:} The line  at 5018.436\,\AA{} is redshifted by about 0.05\,\AA{}
at the phase 0.025, but it is centered at the laboratory
wavelength at the other phases.
However, the observed
profile  is always stronger and broader than the predicted one.

\section{Discussion}
\label{sect:discussion_reproduced}

Among the Herbig~Ae stars, close spectroscopic binaries with orbital periods below 20 days
are very rare \citep{Duchene2015}.
This might be the result of merger events early at the pre-main-sequence stage,
in line with recent observations of magnetic Ap and Bp stars,
suggested to be successors of the magnetic Herbig~Ae/Be stars on the main-sequence.
According to \citet{Ferrario2009}, at least one of the merging stars
must be on the Henyey part of the pre-main-sequence track towards the end
of its contraction to the main sequence. The merger outcome then becomes
observable as a magnetic Herbig~Ae/Be star. This implies that there should be
almost no magnetic star in close Herbig~Ae/Be and Ap/Bp binaries.
Indeed, previous studies of main-sequence binary systems with A- and
late B-type primaries detected only two systems, HD\,98088 and HD\,161701, with
a magnetic Ap star as a component \citep{Babcock1958,Hubrig2014}. 
Therefore, the combination of the determination of the chemical composition
and studies of the magnetic field structure in 
close binary components plays an important
role for understanding the mechanisms that can be responsible
for the generation of magnetic fields.

\begin{figure*}
\centering
\resizebox{0.85\textwidth}{!}{\rotatebox{90}{\includegraphics[18,160][370,826]{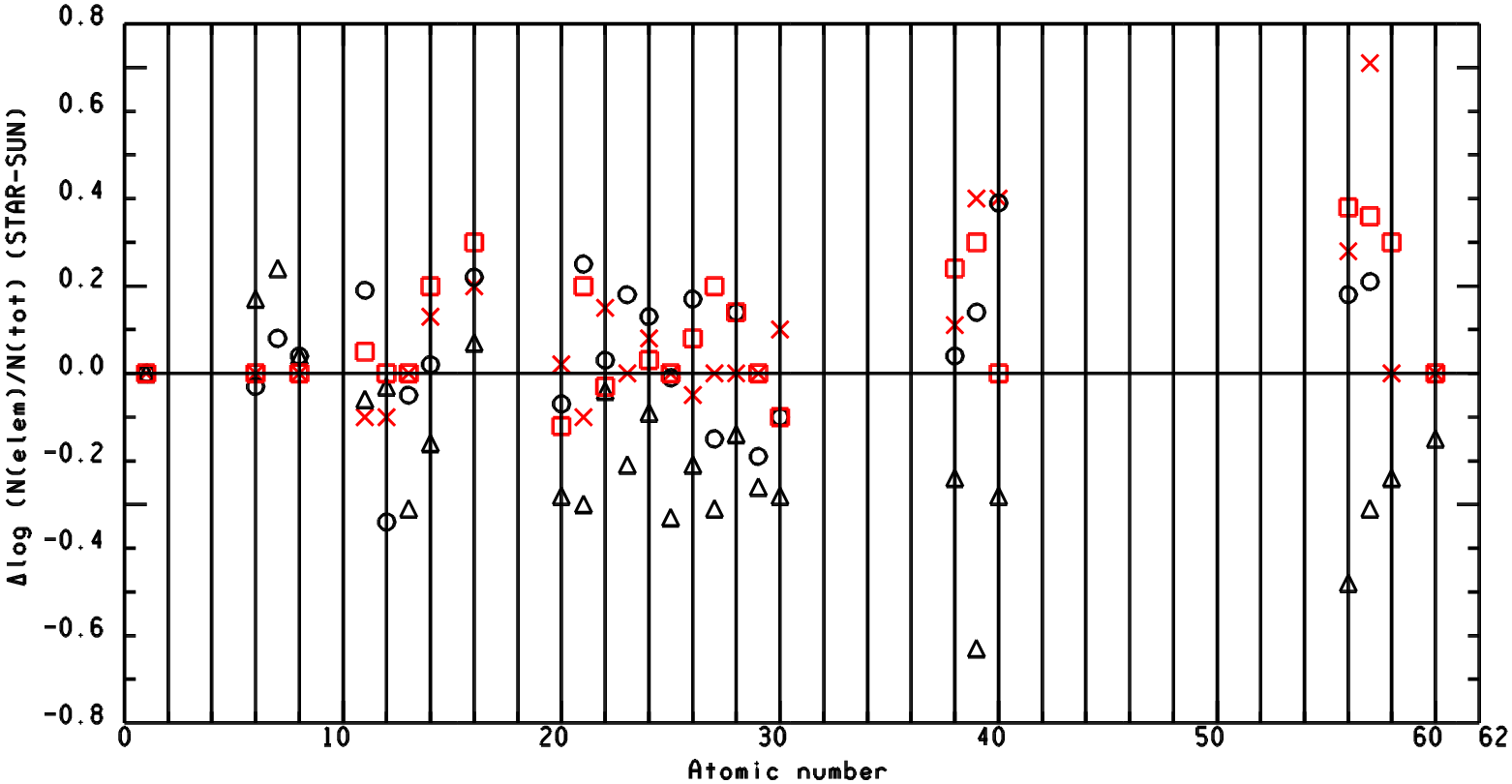}}}
\caption{ Comparison of the abundances relative to the solar ones of
AK\,Sco\,A (red x) and AK\,Sco\,B (red boxes) with the abundances of the
 two magnetic Herbig~Ae stars PDS2 (triangle) and HD\,104237 (open
   circles).
}
\label{fig:f7}
\end{figure*}

Figure~\ref{fig:f7} shows that the abundance patterns of the first (red crosses) and
second (red boxes) components of AK\,Sco have the same trend, although they
are not identical. 
In both stars, most of the elements have solar or nearly solar abundances.
In addition to Li (not included in Fig.~\ref{fig:f7}), exceptions are found for Si, S, Y, Zr, Ba,
and La, which are overabundant. The overabundance in the primary and in the
secondary are correspondingly [2.3] and [2.5] for Li, [0.13] and [0.20] for Si, [0.20] and
[0.30] for S, [0.4] and [0.3] for Y, [0.28] and [0.38] for Ba, [0.71] and
[0.36] for La. Zirconium is overabundant by [0.4] only in the primary, while
it was found solar in the secondary.
In addition to this large abundance  difference for \ion{Zr}{ii},
other remarkable differences are those for \ion{Sc}{ii} (0.30\,dex),
\ion{Ti}{ii} (0.29\,dex), and \ion{La}{ii} (0.35\,dex). Because we estimate
that the errors in the abundance determination are at least of the order
of 0.2\,dex, we prefer to consider the abundance differences
from the two components lower than 0.2\,dex as not conclusive.

In Fig.~\ref{fig:f7} we compare the abundances of the two components of AK\,Sco with the
abundances of the two magnetic Herbig~Ae stars 
HD\,104237 \citep{Cowley2013} and  PDS2 \citep{Cowley2014}.
While the abundance patterns of both components
in the studied SB2 system,
AK\,Sco\,A and AK\,Sco\,B, are similar to the abundance pattern of
the primary in the SB2 system HD\,104237 (open circles), in particular for elements
heavier than strontium, they are rather different from the abundance pattern of
the single Herbig star PDS2 (triangles),
which follows the trend shown by the $\lambda$\,Boo stars.

In conclusion, although AK\,Sco\,B, HD\,104237, and PDS2 are magnetic stars, they do
not have a similar abundance behaviour: the first two stars
display weak Ap/Bp peculiarities, as well as AK\,Sco\,A does, although
no magnetic field was detected up to now for it. In contrast, PDS2
is characterized by $\lambda$\,Boo chemical peculiarities. The two different kinds
of abundance patterns (Ap/Bp and $\lambda$\,Boo) do not seem to be dependent
on temperature, since PDS2 and AK\,Sco have the same $\teff = 6500$\,K.


\section*{Acknowledgements}

Based on observations made with ESO Telescopes at the La Silla Paranal Observatory under programme
IDs 097.C-0277(A) and 099.C-0081(A).
We thank the anonymous referee for the useful comments.








\appendix

\section{The lines analyzed in the spectrum of AK\,Sco}
\label{sect:lines}

In Table~\ref{tab:t3}, the lines analysed in the spectrum of AK\,Sco for abundance purposes are listed.
Successive columns display wavelength, $\log\,gf$ value,
the source for the $\log\,gf$ value, the lower excitation
potential in cm$^{-1}$, the abundance from the primary (Star A), remarks about the given
line in the primary, the abundance from the secondary (Star B), and remarks about the given
line in the secondary. If the abundance was very uncertain, owing to blends or
other causes, no abundance is indicated.
For each examined ion the line at the top gives the average abundance for the ion
under consideration
and the solar abundance. If the element is present in two ionization states and different
abundances from the two ions were derived, a line after the entries for the two ions gives the average
abundance obtained from all lines of that element. This is the case for \ion{Si}{i}, \ion{Si}{ii},
\ion{Ti}{i}, \ion{Ti}{ii}, \ion{Cr}{i}, \ion{Cr}{ii}, \ion{Fe}{i}, \ion{Fe}{ii}, \ion{Sr}{i}, and
\ion{Sr}{ii}.

\vskip1mm

{\bf Notes to Table~\ref{tab:t3}:}

($^{a}$)
The index ``h'' added after the wavelength value indicates that hyperfine structure was considered for that line.

($^{b}$)
a ``K:'' at the beginning of the $\log gf$ source (Col.~3) indicates that the $\log gf$ value
was taken from the Kurucz line list available at
http://kurucz.harvard.edu/ linelists/gfnew/gfall08oct17.dat
and its reference from
http://kurucz.harvard.edu/linelists/gfnew/gfall.ref;
NIST5: NIST database -- http://www.nist.gov/PHYSRefData/ ASD/lines-form.html \citep{Kramida2018};
ALD: \citet{Aldenius2009};
FMW: \citet{Fuhr1988};
GARZ: \citet{Garz1973}; 
K88: \cite{Kurucz1993};
Ljun: \cite{Ljung2006};
PTP: \cite{Pickering2001};
RU1: \cite{RaassenUylings1998a};
RU2: \cite{RaassenUylings1998b};
SUN: astrophysical $\log gf$ value derived by fitting the observed solar profiles.

($^{c}$) Significance of the notes in Columns~6 and 8:
bl A -- line blended with other lines in the primary;
bl B -- line blended with other lines in the secondary;
bl AB -- line blended with other lines in both primary and secondary;
bl unk -- line blended with an unidentified line;
bl -- line blended for other reasons;
single -- line unblended;
almost single -- line blended with a very minor component;
redsh -- line redshifted;
core -- the observed line core is weaker than the computed one;
TSC -- the line is computed too strong;
red comp -- line affected by a red component;
cont -- line affected by continuum.

\begin{table*}
\caption[ ]{Abundances for the two components of AK\,Sco from the ATLAS9 model with parameters
  $\teff=6500$\,K, $\logg=4.5$, $v_{\rm turb}=1.0$ and 2.0\,km\,s$^{-1}$ for the primary and the secondary, respectively.}
\label{tab:t3}
\centering
\begin{tabular}{lclr @{.} lcl|cl}
\hline\noalign{\smallskip}
\multicolumn{1}{c}{$\lambda^{a}$ [\AA{}]} &
\multicolumn{1}{c}{$\log gf$}&
\multicolumn{1}{c}{Ref.$^{b}$}&
\multicolumn{2}{c}{$\chi_{\rm low}$}&
\multicolumn{1}{c}{$\log \frac{N_{\rm Z}}{N_{\rm tot(A})}$}&
\multicolumn{1}{l}{Notes$^{c}$}&
\multicolumn{1}{c}{$\log \frac{N_{\rm Z}}{N_{\rm tot(B})}$}&
\multicolumn{1}{l}{Notes$^{c}$} \\
\hline\noalign{\smallskip}
\multicolumn{1}{c}{\ion{Li}{i}}   & & & \multicolumn{2}{c}{} & $-$8.70 & $-$10.99 (Sun)  &$-$8.50 &$-$10.99 (Sun) \\
\hline\noalign{\smallskip}
6707.76   & $-$0.002 & NIST5  &     0&000 & $-$8.70 & bl \ion{Li}{i} & $-$8.50 & bl \ion{Li}{i} \\
6707.91   & $-$0.303 & NIST5  &     0&000 & $-$8.70 & bl \ion{Li}{i} & $-$8.50 & bl \ion{Li}{i} \\
\hline\noalign{\smallskip}
\multicolumn{1}{c}{\ion{C}{i}}    & & & \multicolumn{2}{c}{} & $-$3.61 &$-$3.61 (Sun) &$-$3.61 &$-$3.61 (Sun)) \\
\hline\noalign{\smallskip}
4932.050  & $-$1.658 & NIST5  & 61981&818 &  & bl AB  &   & bl AB  \\
5052.144  & $-$1.303 & NIST5  & 61981&818 & $-$3.61 & single  &      & bl unk \\
5380.325  & $-$1.616 & NIST5  & 61981&818 &  & bl B    &  &bl A \\
6587.620  & $-$1.003 & NIST5  & 68856&338 &  & bl AB  & $-$3.61 & single\\
\hline\noalign{\smallskip}
\multicolumn{1}{c}{\ion{O}{i}}    & & & \multicolumn{2}{c}{} & $-$3.35 &$-$3.35 (Sun) &$-$3.35 &$-$3.35 (Sun) \\
\hline\noalign{\smallskip}
6155.961  & $-$1.363 & NIST5  & 86625&757 & $-$3.35 & weak, bl AB & $-$3.35  & weak, bl AB \\
6155.971  & $-$1.011 & NIST5  & 86625&757 & $-$3.35 & weak, bl AB &          & weak, bl AB\\
6155.989  & $-$1.120 & NIST5  & 86625&757 & $-$3.35 & weak, bl AB &          & weak, bl AB\\
6156.737  & $-$1.487 & NIST5  & 86627&778 & $-$3.35 & weak, bl AB & $-$3.35  & weak, bl AB\\
6156.755  & $-$0.898 & NIST5  & 86627&778 & $-$3.35 & weak, bl AB & $-$3.35  & weak, bl AB\\
6156.778  & $-$0.694 & NIST5  & 86627&778 & $-$3.35 & weak, bl AB & $-$3.35  & weak, bl AB\\
6158.149  & $-$1.841 & NIST5  & 86631&454 & $-$3.35 & weak, bl B   & $-$3.35: & weak, bl AB \\
6158.172  & $-$0.996 & NIST5  & 86631&454 & $-$3.35 & weak, bl B   & $-$3.35: & weak, bl AB\\
6158.187  & $-$0.409 & NIST5  & 86631&454 & $-$3.35 & weak, bl B   & $-$3.35: & weak, bl AB\\
\hline\noalign{\smallskip}
\multicolumn{1}{c}{\ion{Na}{i}}   & & & \multicolumn{2}{c}{} & $-$5.93 &$-$5.83 (Sun) &$-5.78\pm0.05$ &$-$5.83 (Sun) \\
\hline\noalign{\smallskip}
4982.813h & $-$0.962 & NIST5  & 16973&366 &   & bl B &  & bl A\\
5682.633h & $-$0.706 & NIST5  & 16956&172 & $-$5.93  & single & $-$5.73 & bl \\
5688.193h & $-$1.406 & NIST5  & 16973&368 &   & bl B & $-$5.83 & almost single \\
5688.295h & $-$0.452 & NIST5  & 16973&368 &   & bl B & $-$5.83 & almost single \\
5889.950h & $+$0.108 & NIST5  &     0&000 &   &complex structure &         & complex structure \\
5895.924h & $-$0.194 & NIST5  &     0&000 &   &complex structure &         & complex structure \\
6154.225h & $-$1.547 & NIST5  & 16956&170 & $-$5.93 & single&  & bl A \\
6160.747h & $-$1.246 & NIST5  & 16973&366 &  & bl B &  &  bl A\\ 
\hline\noalign{\smallskip}
\multicolumn{1}{c}{\ion{Mg}{i}}   & & & \multicolumn{2}{c}{} & $-$4.55 &$-$4.45 (Sun)  &$-$4.45 &$-$4.45 (Sun)\\
\hline\noalign{\smallskip}
4057.505  & $-$0.900 & NIST5  & 35051&264 &   & bl AB  & & bl AB \\ 
4167.271  & $-$0.745 & NIST5  & 35051&264 & $-$4.55 & single  & $-$4.45 & single \\
4702.991  & $-$0.440 & NIST5  & 35051&264 &          & bl A, core TSC &  &bl B\\
5167.321  & $-$0.870 & NIST5  & 21850&405 & & bl AB, core TSC, redsh 1.5\,km\,s$^{-1}$ &  & bl AB\\
5172.684  & $-$0.393 & NIST5  & 21870&464 & &bl AB, core TSC, redsh 1.5\,km\,s$^{-1}$&     & bl AB  \\ 
5183.604  & $-$0.167 & NIST5  & 21911&178 &    & bl A, core TSC &  & bl AB   \\
5528.405  & $-$0.498 & NIST5  & 35051&264 & $-$4.55  & bl B, core TSC & $-$4.45& single  \\
5711.095  & $-$1.724 & NIST5  & 35051&264 &  &bl B & $-$4.46& single\\
\hline\noalign{\smallskip}
\multicolumn{1}{c}{\ion{Mg}{ii}}  & & & \multicolumn{2}{c}{} & $-$4.55 &$-$4.45 (Sun)&$-$4.45 &$-$4.45 (Sun)\\
\hline\noalign{\smallskip}
4481.126  & $+$0.749 & NIST5  & 71490&190 & $-$4.55 & bl B & $-$4.45 & bl A \\ 
4481.150  & $-$0.553 & NIST5  & 71490&190 & $-$4.55 & bl B & $-$4.45 &      \\
4481.325  & $+$0.594 & NIST5  & 71491&063 & $-$4.55 & bl B & $-$4.45 &      \\ 
\hline\noalign{\smallskip}
\multicolumn{1}{c}{\ion{Al}{i}}   & & & \multicolumn{2}{c}{} & $-$5.61 &$-$5.61 (Sun) &$-$5.61 &$-$5.61 (Sun) \\
\hline\noalign{\smallskip}
3944.006h & $-$0.638 & NIST5  &     0&000 & $-$5.61&bl B, redsh 3\,km\,s$^{-1}$ & $-$5.61 & bl AB \\
3961.520h & $-$0.336 & NIST5  &   112&061 & $-$5.61 &bl B & $-$5.61  & bl A \\
6696.018h & $-$1.569 & NIST5  & 25347&756 &  & bl AB &       & bl AB \\
6698.667h & $-$1.870 & NIST5  & 25347&756 &  & single, bad cont ? &  & single, bad cont ?\\
\hline
\end{tabular}
\end{table*}

\setcounter{table}{0}

\begin{table*}
\caption[ ]{cont.}
\centering
\begin{tabular}{lclr @{.} lcl|cl}
\hline\noalign{\smallskip}
\multicolumn{1}{c}{$\lambda^{a}$ [\AA{}]} &
\multicolumn{1}{c}{$\log gf$}&
\multicolumn{1}{c}{Ref.$^{b}$}&
\multicolumn{2}{c}{$\chi_{\rm low}$}&
\multicolumn{1}{c}{$\log \frac{N_{\rm Z}}{N_{\rm tot(A})}$}&
\multicolumn{1}{l}{Notes$^{c}$}&
\multicolumn{1}{c}{$\log \frac{N_{\rm Z}}{N_{\rm tot(B})}$}&
\multicolumn{1}{l}{Notes$^{c}$} \\
\hline\noalign{\smallskip}
\multicolumn{1}{c}{\ion{Si}{i}}   & & & \multicolumn{2}{c}{} & $-4.40\pm0.07$ &$-$4.53 (Sun) &$-4.31\pm0.06$ &$-$4.53 (Sun) \\
\hline\noalign{\smallskip}
5645.613  & $-$2.140 & GARZ   & 39760&285 &  & bl B &  $-$4.33 & almost single\\
5665.555  & $-$2.040 & NIST5  & 39683&163 &  & bl B   &     & bl A\\
5690.425  & $-$1.870 & NIST5  & 39760&285 & $-$4.33 & single&  $-$4.23 & single \\
5708.400  & $-$1.470 & NIST5  & 39955&053 &  $-$4.28 &almost single  &  & bl AB  \\
5772.146  & $-$1.750 & NIST5  & 40991&884 &  $-$4.28 & single &    $-$4.23 & almost single\\
5948.541  & $-$1.231 & NIST5  & 40991&884 & $-$4.43 & single &  & bl with telluric \\
6091.919  & $-$1.250 & SUN    & 47351&554 &         & bl AB &  & bl A \\
6125.021  & $-$1.565 & SUN    & 45276&188 &          & bl AB &  & bl A\\
6142.483  & $-$1.480 & SUN    & 45321&848 & $-$4.43 & single &  $-$4.33 & single\\
6145.016  & $-$1.411 & SUN    & 45293&629 & $-$4.43 & single &  $-$4.33 & single  \\
6155.134  & $-$0.855 & SUN    & 45321&848 & $-$4.43 & single &  & almost single, bad cont\\
6237.319  & $-$1.070 & SUN    & 45276&188 & $-$4.48 & single, redsh &  $-$4.43 & single  \\
6243.815  & $-$1.244 & K:K07  & 45293&629 & $-$4.48 & single &   & bl A\\
6244.466  & $-$1.291 & SUN    & 45293&629 & $-$4.38 & single &  & bl A  \\  
6414.980  & $-$1.036 & K:K07  & 47351&554 & $-$4.48 & single &  & bl A\\
6721.848  & $-$1.140 & SUN    & 47284&061 &  $-$4.33 & single & $-$4.33 & single\\
\hline\noalign{\smallskip}
\multicolumn{1}{c}{\ion{Si}{ii}}&  & & \multicolumn{2}{c}{} & $-$4.40 &$-$4.53 (Sun)  &$-$4.40 &$-$4.53 (Sun) \\
\hline\noalign{\smallskip}
5055.984  & $+$0.523 & NIST5  & 81251&320 &    & bl AB & & bl AB, telluric \\
6347.109  & $+$0.149 & NIST5  & 65500&470 &   $-$4.40 & single & $-$4.40& single \\
6371.359  & $-$0.082 & NIST5  & 65500&470 &           & bl B &  $-$4.40& single \\ 
\hline\noalign{\smallskip}
\multicolumn{1}{c}{Si(tot)} & & & \multicolumn{2}{c}{} & $-4.40\pm0.07$ &$-$4.53 (Sun) & $-4.33\pm0.07$ & $-$4.53 (Sun)  \\
\hline
\hline\noalign{\smallskip}
\multicolumn{1}{c}{\ion{S}{i}}&  & & \multicolumn{2}{c}{} & $-4.72\pm0.10$ &$-$4.92 (Sun)  &$-4.62\pm0.10$ &$-$4.92 (Sun) \\
\hline\noalign{\smallskip}
6046.038  & $-$0.959 & NIST5  & 63457&142 &    & bl B &  $-$4.52 & single \\
6052.594  & $-$1.258 & NIST5  & 63475&051 &   $-$4.82 & single & & bl B\\
6743.552  & $-$1.065 & NIST5  & 63446&065 &   & bl AB &  & bl A\\
6748.578  & $-$0.843 & NIST5  & 63457&142 &    & bl unk &  & bl unk\\
6748.792  & $-$0.638 & NIST5  & 63457&142 &    & bl unk & &bl A, unk\\
6756.961  & $-$0.937 & NIST5  & 63477&051 &    $-$4.62  & bl \ion{S}{i}& $-$4.72 & bl \ion{S}{i} \\
6757.153  & $-$0.351 & NIST5  & 63475&051 &      $-$4.62  & bl \ion{S}{i}& $-$4.72 & bl \ion{S}{i} \\
\hline\noalign{\smallskip}
\multicolumn{1}{c}{\ion{Ca}{i}}&  & & \multicolumn{2}{c}{} & $-5.70\pm0.10$ &$-$5.72 (Sun)  &$-5.84\pm0.08$ &$-$5.72 (Sun)\\
\hline\noalign{\smallskip}
4226.728  & $+$0.244 & NIST5  &     0&000 &           &   bl AB    &  & bl AB  \\
4283.011  & $-$0.224 & NIST5  & 15210&063 &       &   bl AB &  & bl  AB  \\
4318.652  & $-$0.208 & NIST5  & 15315&943 &       &   bl A & $-$5.65  & single   \\
4425.437  & $-$0.358 & NIST5  & 15157&901 & $-$5.72 & almost single & $-$5.72  & almost single \\
4434.957  & $-$0.010 & NIST5  & 15210&063 &         & bl AB  &                & bl B  \\
4435.679  & $-$0.519 & NIST5  & 15210&063 &         & bl unk ? &               & bl A \\
4578.551  & $-$0.558 & NIST5  & 20335&360 & $-$5.72 & single &                 & bl unk ?  \\
4585.865  & $-$0.187 & NIST5  & 20371&000 &       &bl B &                    & bl A\\
4685.268  & $-$0.88  & NIST5  & 44989&830 &$-$5.57  &single &                & bl A\\
5512.980  & $-$0.300 & NIST5  & 23652&304 &  $-$5.82 & almost single &         & bl A\\
5581.965  & $-$0.517 & K:K07  & 20349&260 &  $-$5.57 & almost single & $-$5.62 & almost single\\
5590.114  & $-$0.547 & K:K07  & 20335&360 &          & bl B &  $-$5.62 & almost single    \\
5598.480  & $-$0.091 & K:K07  & 20335&360 &          & bl A &          & bl B\\
5867.562  & $-$1.600 & SUN    & 23652&304 &          & bl B &  $-$5.62 & single \\
6102.723  & $-$0.790 & NIST5  & 15157&901 & $-$5.57 & single & $-$5.62 & single  \\ 
6122.217  & $-$0.315 & NIST5  & 15210&063 & $-$5.65 & single &  $-$5.55 & single \\
6162.173  & $-$0.170 & ALD    & 15315&943 & $-$5.65 & almost single &    & bl AB    \\
6169.042  & $-$0.54  & NIST5  & 20349&268 & $-$5.77 & almost single &  & bl A\\                        
6169.563  & $-$0.27  & NIST5  & 20371&000 & $-$5.72 & almost single &  & bl A \\
6439.075  & $+$0.47  & NIST5  & 20371&000 & $-$5.70 & single        & $-$5.60 & single \\
6449.808  & $-$0.55  & NIST5  & 20335&360 & $-$5.70 & almost single & $-$5.70 & almost single \\
6471.662  & $-$0.59  & NIST5  & 20371&000 & $-$5.70 & single & $-$5.70 &single, bl telluric \\
6499.650  & $-$0.59  & NIST5  & 20371&000 & $-$5.95 & single & $-$5.80 & single\\
6717.681  & $-$0.61  & NIST5  & 21849&634 & $-$5.70 & almost single & $-$5.45 & almost single, \\
          &          &        & \multicolumn{2}{c}{} & &            &         & redsh 1.5\,km\,s$^{-1}$, red comp ?   \\
\hline
\end{tabular}
\end{table*}

\setcounter{table}{0}

\begin{table*}
\caption[ ]{cont.}
\centering
\begin{tabular}{lclr @{.} lcl|cl}
\hline\noalign{\smallskip}
\multicolumn{1}{c}{$\lambda^{a}$ [\AA{}]} &
\multicolumn{1}{c}{$\log gf$}&
\multicolumn{1}{c}{Ref.$^{b}$}&
\multicolumn{2}{c}{$\chi_{\rm low}$}&
\multicolumn{1}{c}{$\log \frac{N_{\rm Z}}{N_{\rm tot(A})}$}&
\multicolumn{1}{l}{Notes$^{c}$}&
\multicolumn{1}{c}{$\log \frac{N_{\rm Z}}{N_{\rm tot(B})}$}&
\multicolumn{1}{l}{Notes$^{c}$} \\
\hline\noalign{\smallskip}
\multicolumn{1}{c}{\ion{Ca}{ii}}& & & \multicolumn{2}{c}{} & &$-$5.72 (Sun) & & $-$5.72(Sun) \\
\hline\noalign{\smallskip}
3933.664  & 0.111    & K:K10  &     0&000 &  & core not fitted &         & not fitted    \\
3968.469  & $-$0.194 & K:K10  &     0&000 &  & core not fitted &          & not fitted       \\
\hline\noalign{\smallskip}
\multicolumn{1}{c}{\ion{Sc}{ii}}& & & \multicolumn{2}{c}{} & $-$8.98 & $-$8.88 (Sun) &$-$8.68  &$-$8.88 (Sun) \\
\hline\noalign{\smallskip}
4246.822h & $+$0.242 & NIST5  &  2540&950 &          & bl, red comp ? &   & bl, red comp ?\\ 
4314.083h & $-$0.096 & NIST5  &  4987&790 &          & bl AB  &     & bl AB \\
4320.732h & $-$0.252 & NIST5  &  4883&570 &          & bl A    &     & bl AB \\
4400.389h & $-$0.536 & NIST5  &  4883&570 & $-$8.98 & almost single &     & bl A \\ 
4670.407h & $-$0.576 & NIST5  & 10944&560 &         & bl AB  &      & bl AB \\
5031.021  & $-$0.399 & NIST5  & 10944&560 &          & bl AB & $-$8.68 & almost single\\         
5239.813  & $-$0.765 & NIST5  & 11736&360 & $-$8.98 & almost single &  $-$8.68& almost single \\         
5526.790h & $+$0.025 & NIST5  & 14261&250 &         & bl B, red comp ? &   & bl A \\         
6604.601h & $-$1.309 & NIST5  & 10944&560 & $-$8.98 & single &   & bl B   \\         
\hline\noalign{\smallskip}
\multicolumn{1}{c}{\ion{Ti}{i}}& & & \multicolumn{2}{c}{} & $-7.145\pm0.04$ & $-$7.11 (Sun) &$-$7.11 &$-$7.11 (Sun) \\
\hline\noalign{\smallskip}
4286.004  & $-$0.350 & K:LGWS &  6661&006 &  $-$7.11 & single &       & bl \\
4512.734  & $-$0.424 & K:BMPS &  6742&756 &  $-$7.21 & almost single &    & spike \\ 
4981.731  & $+$0.560 & K:BMPS &  6842&962 &          & bl B &     & bl A\\          
4999.503  & $+$0.306 & K:BMPS &  6661&004 &          & bl B &     & bl A\\
5035.903  & $+$0.220 & K:LGWS & 11776&812 &           & bl A &      & bl B\\
5039.957h & $-$1.074 & K:BMP  &   170&13  &           & bl B &      &no fit, bl A\\
5866.451h & $-$0.784 & K:BMPS &  8602&344 &  $-$7.11 & single &     & bl A\\
\hline\noalign{\smallskip}
\multicolumn{1}{c}{\ion{Ti}{ii}}& & & \multicolumn{2}{c}{} & $-6.88\pm0.09$ &$-$7.11 (Sun) &$-7.14\pm0.09$&$-$7.11 (Sun)\\
\hline\noalign{\smallskip}
4025.129  & $-$2.110 & K:WLSC &  4897&650 &            & bl AB  &       & bl B  \\         
4287.873h & $-$1.790 & PTP    &  8710&440 &             & bl AB &        & bl AB \\         
4290.215h & $-$0.870 & K:WLSC &  9395&710 &            & bl AB &        & bl AB\\         
4300.042h & $-$0.460 & K:WLSC &  9518&060 &            & bl AB &       & bl AB\\         
4312.860h & $-$1.120 & K:WLSC &  9518&152 & $-$6.81    & single &        & bl A \\ 
4398.289  & $-$2.650 & K:PTP  &  9872&899 &            & bl A &          & bl AB \\
4399.765h & $-$1.200 & K:WLSC &  9975&999 &            & bl A  &         & bl A \\ 
4411.073h & $-$0.650 & K:WLSC & 24961&030 &            & bl A  &        & bl B \\         
4411.930  & $-$2.620 & K:WLSC &  9872&73  &            & bl A &         & bl B \\         
4417.713h & $-$1.167 & K:K16  &  9395&802 &            & bl B &         & bl A \\
4421.938  & $-$1.640 & K:WLSC & 16625&11  &             & bl A &         & bl AB \\         
4441.728  & $-$2.333 & K:K16  &  9518&06  & $-$6.81 & almost single &      & bl A \\          
4443.801h & $-$0.710 & K:WLSC &  8710&44  &       & red comp ?, not in syn1  &       & bl unk ?\\         
4450.482h & $-$1.520 & K:WLSC &  8744&25  &       & bl AB &             & bl AB \\         
4468.492h & $-$0.600 & K:BHNG &  9118&26  &     $-$6.81 & single &        & bl A\\          
4488.325h & $-$0.500 & K:WLSC & 25192&79  &           & bl A &          & bl AB  \\         
4501.270h & $-$0.770 & K:WLSC &  8997&71  &            & bl AB &        & bl AB \\         
4518.332  & $-$2.560 & K:WLSC &  8710&440 &             & bl A&           & bl AB  \\         
4533.960h & $-$0.577 & K:K16  &  9975&92  &             & bl AB &        & bl AB\\         
4563.758h & $-$0.795 & K:K16  & 10024&73  &             & bl AB &        & bl AB \\         
4571.971h & $-$0.310 & K:WLSC & 12676&97  &            & bl AB, red comp ? &  & bl B, red comp ? \\         
4708.663  & $-$2.350 & K:WLSC &  9975&92  &  & bl B &  & bl A  \\         
4779.979  & $-$1.248 & K:K16  & 16515&86  &  $-$6.81 & single & $-$7.01 & single\\         
4798.531  & $-$2.660 & K:WLSC &  8710&44  &         & bl A &            & bl AB \\         
5010.209  & $-$1.350 & K:WLSC & 25192&79  &         & bl A &          & bl AB \\         
5211.530  & $-$1.410 & K:WLSC & 20891&787 &         & bl B & $-$7.21 & almost single \\         
5381.022  & $-$1.970 & K:WLSC & 12628&73  &         & bl B &         & bl A \\         
5418.768  & $-$2.130 & K:WLSC & 12758&11  & $-$7.01& single &        & bl A \\         
5490.693  & $-$2.663 & K:K16  & 12628&834 &      & bl A   &        & bl A \\         
6491.566  & $-$1.942 & K:K16  & 16625&110 & $-$7.01: & bl telluric &  $-$7.21  & single \\         
\hline\noalign{\smallskip}
\multicolumn{1}{c}{Ti(tot)} & & & \multicolumn{2}{c}{} & $-6.96\pm0.15$&$-$7.11 (Sun) & $-7.14\pm0.09$ & $-$7.11 (Sun)  \\
\hline
\hline\noalign{\smallskip}
\end{tabular}
\end{table*}

\setcounter{table}{0}

\begin{table*}
\caption[ ]{cont.}
\centering
\begin{tabular}{lclr @{.} lcl|cl}
\hline\noalign{\smallskip}
\multicolumn{1}{c}{$\lambda^{a}$ [\AA{}]} &
\multicolumn{1}{c}{$\log gf$}&
\multicolumn{1}{c}{Ref.$^{b}$}&
\multicolumn{2}{c}{$\chi_{\rm low}$}&
\multicolumn{1}{c}{$\log \frac{N_{\rm Z}}{N_{\rm tot(A})}$}&
\multicolumn{1}{l}{Notes$^{c}$}&
\multicolumn{1}{c}{$\log \frac{N_{\rm Z}}{N_{\rm tot(B})}$}&
\multicolumn{1}{l}{Notes$^{c}$}\\
\hline\noalign{\smallskip}
\multicolumn{1}{c}{\ion{V}{i}}& & & \multicolumn{2}{c}{} &$-$8.15 &$-$8.15(Sun) & --- &$-$8.15 (Sun) \\
\hline\noalign{\smallskip}
4379.230h & $+$0.580 & NIST5  & 2424&780 & $-$8.15 & single &  & bl A\\
6090.208h & $-$0.067 & NIST5  & 8715&760 & $-$8.15 & single &  & bl A\\
6119.527h & $-$0.350 & NIST5  & 8578&530 &         & bl B &    & bl AB, weak\\
\hline\noalign{\smallskip}
\multicolumn{1}{c}{\ion{V}{ii}}& & & \multicolumn{2}{c}{} &\multicolumn{1}{c}{---} & &\multicolumn{1}{c}{---} &$-$8.15 (Sun) \\
\hline\noalign{\smallskip}
4002.928  & $-$1.440 & K:WLDS & 11514&760 &             & bl AB &             & bl AB \\
4023.377  & $-$0.610 & K:WLDS & 14556&090 &             & bl B   &               & bl A \\
4183.428  & $-$1.060 & K:WLDS & 16533&00  &              & bl AB &             & bl A \\
4312.348  & $-$1.495 & K:BGFM & 13490&883 &             & bl AB &             & bl B\\ 
\hline\noalign{\smallskip}
\multicolumn{1}{c}{\ion{Cr}{i}}& & & \multicolumn{2}{c}{} &$-6.41\pm0.12$ &$-$6.42 (Sun) &$-$6.42&$-$6.42 (Sun) \\
\hline\noalign{\smallskip}
4254.336  & $-$0.108 & K:BMP  &     0&000 &  & bl AB &  & bl AB\\ 
4274.797  & $-$0.231 & K:BMP  &     0&000 &  & bl AB &   & bl AB\\ 
4616.124  & $-$1.204 & K:BMP  &  7927&441 &  $-$6.42 & single &  & bl A\\  
4626.173  & $-$1.340 & K:BMP  &  7810&820 & & bl B&  & bl A \\ 
4652.157  & $-$1.035 & K:SLS  &  8095&187 & & bl B &$-$6.42  & almost single \\ 
4718.420  & $+$0.240 & K:SLS  & 25771&420 &  $-$6.42 & almost single, redsh ? & $-$6.42 & almost single, redsh ? \\ 
5204.511  & $-$0.198 & K:BMP  &  7593&160 &    & bl A&  & bl AB \\ 
5208.425  & $+$0.172 & K:BMP  &  7593&148 &  & bl A, flat core &   & bl B\\ 
5409.784  & $-$0.715 & K:BMP  &  8307&575 &  $-$6.22 & almost single &  & bl AB\\
5787.919  & $-$0.083 & K:BBMP & 26796&266 & $-$6.57 & single & $-$6.42 & single\\
5790.957  & $+$0.250 & K:K16  & 25787&964 &  & bl A, flat core &  & bl B\\
\hline\noalign{\smallskip}
\multicolumn{1}{c}{\ion{Cr}{ii}}& & & \multicolumn{2}{c}{} &$-6.22\pm0.2$ &$-$6.42 (Sun) &$-6.35\pm0.04$ & $-$6.42 (Sun)\\
\hline\noalign{\smallskip}
4242.366  & $-$1.352 & K:K16  & 31219&350 &        & bl AB &              & bl AB\\ 
4252.625  & $-$2.053 & K:K16  & 31117&390 &        & bl B   &              & bl A\\ 
4558.644  & $-$0.444 & K:K16  & 32854&31  & $-$6.42 & bl \ion{Cr}{ii} & $-$6.32 & almost single \\ 
4558.788  & $-$2.741 & RU1    & 32854&94  &  $-$6.42 & bl \ion{Cr}{ii}  &  $-$6.32 &   \\ 
4588.198  & $-$0.826 & K:NLLN & 32836&68  &       &bl B &  $-$6.32 & single\\ 
4616.624  & $-$1.575 & RU1    & 32844&76  &       & bl AB &  & bl B\\ 
4618.807  & $-$0.996 & K:NLLN & 32854&95  &  & bl A&    & bl AB\\ 
4634.070  & $-$0.980 & K:NLLN & 32844&76  &  & bl B &  & bl A\\
4824.131  & $-$1.085 & RU1    & 31219&332 &  & bl A &  & bl AB\\   
5237.322  & $-$1.160 & NIST5  & 32854&31  &  & bl AB & $-$6.42 & almost single\\ 
5502.085  & $-$2.048 & K:K16  & 33618&91  & $-$6.02  & single &  & bl A\\ 
\hline\noalign{\smallskip}
\multicolumn{1}{c}{Cr(tot)} & & & \multicolumn{2}{c}{} &$-6.34\pm0.17$ &$-$6.42 (Sun)& $-6.39\pm0.02$ &$-$5.42 (Sun)  \\
\hline
\hline\noalign{\smallskip}
\multicolumn{1}{c}{\ion{Mn}{i}}& & & \multicolumn{2}{c}{} &$-$6.62 &$-$6.62 (Sun) &$-$6.62 & $-$6.62 (Sun)\\
\hline\noalign{\smallskip}
4033.062h & $-$0.647 & K:DLSS &     0&000 &       & bl AB &  & bl B \\
4034.483h & $-$0.843 & K:DLSS &     0&000 & $-$6.62& single &  &  bl A \\
4041.355h & $+$0.281 & K:DLSS & 17052&29  &  & bl A &  &  bl B \\
4754.042h & $-$0.080 & K:DLSS & 23719&52  & $-$6.62 & almost single &  $-$6.62 & almost single \\
4783.427h & $+$0.044 & K:DLSS & 18531&5   & $-$6.62 & almost single &  $-$6.62 & single  \\
4823.524h & $+$0.136 & K:DLSS & 18705&2   &         & bl A &        & bl B  \\
5377.596h & $-$0.166 & K:K07  & 31001&153 &        & bl B &        & bl A \\
5537.765h & $-$2.017 & K88    & 17637&156 &         & bl AB  &      & bl AB\\
6013.510h & $-$0.354 & K:DLSS & 24779&331 &       & red comp ?, &  & red comp ?, \\
          &          &        & \multicolumn{2}{c}{} & & only $\phi=0.090$ & & only $\phi=0.090$ \\
\hline
\end{tabular}
\end{table*}

\setcounter{table}{0}

\begin{table*}
\caption[ ]{cont.}
\centering
\begin{tabular}{lclr @{.} lcl|cl}
\hline\noalign{\smallskip}
\multicolumn{1}{c}{$\lambda^{a}$ [\AA{}]} &
\multicolumn{1}{c}{$\log gf$}&
\multicolumn{1}{c}{Ref.$^{b}$}&
\multicolumn{2}{c}{$\chi_{\rm low}$}&
\multicolumn{1}{c}{$\log \frac{N_{\rm Z}}{N_{\rm tot(A})}$}&
\multicolumn{1}{l}{Notes$^{c}$}&
\multicolumn{1}{c}{$\log \frac{N_{\rm Z}}{N_{\rm tot(B})}$}&
\multicolumn{1}{l}{Notes$^{c}$}\\
\hline\noalign{\smallskip}
\multicolumn{1}{c}{\ion{Fe}{i}} & & & \multicolumn{2}{c}{} & $-4.66\pm0.15$ &$-$4.57 (Sun) &$-4.51\pm0.17$ & $-$4.57 (Sun)  \\
\hline\noalign{\smallskip}
4071.738  & $-$0.022 & K:FW   & 12968&554 &        & bl AB &  & bl AB\\
4383.544  & $+$0.200 & K:FW   & 11976&239 &        & bl A  &  & bl AB\\
4736.773  & $-$0.67  & K:DRLP & 25899&899 &  &bl B & $-$4.57 &single \\
4890.755  & $-$0.38  & K:RDLB & 23192&500 &  & bl A &  & bl B\\
4891.492  & $-$0.112 & K:FW   & 22996&674 & $-$4.67 & single & & bl A   \\
4903.310  & $-$0.89  & K:RDLB & 23244&838 &  & bl A&  & bl AB \\
4907.732  & $-$1.70  & K:DRLP & 27666&348 &  & bl A& & bl AB \\
4909.383  & $-$1.325 & K:K17  & 31686&351 &    & bl B &  & bl A \\
4917.230  & $-$1.66  & K:FW   & 33801&572 & $-$4.57& single &  & bl A \\
4918.954  & $-$0.602 & K:K17  & 33507&123 & $-$4.57 & bl \ion{Fe}{i} &  & bl AB \\
4918.994  & $-$0.342 & K:FW   & 23110&939 & $-$4.57 & bl \ion{Fe}{i} & $-$4.57& bl AB \\
4920.502  & $+$0.07  & K:RDLB & 22845&869 &        & bl AB, core &      & very bl AB \\
4946.387  & $-$1.11  & K:RDLB & 27166&820 &  & bl AB &   &  bl A \\
4950.105  & $-$1.50  & K:DRLP & 27559&583 &  & single, bad cont ? &  $-$4.37 & single \\
4957.596  & $+$0.233 & K:FW   & 22650&416 &  &bl A, core &   & bl AB \\
4962.571  & $-$1.182 & K:FW   & 33695&397 &  &bl B &   $-$4.27 & single  \\
4969.917  & $-$0.588 & K:K17  & 34017&103 &  &bl B &  & bl A  \\
4970.496  & $-$1.74  & K:FW   & 29320&025 &  &bl A &  & bl B  \\
4991.268  & $-$0.368 & K:K17  & 33801&572 &  &bl AB, core &  & bl B  \\
4994.129  & $-$3.080 & K:FW   &  7376&764 &   &bl B&   & bl B  \\
5007.274  & $-$0.198 & K:K17  & 33095&941 &  &bl AB, core  &  & bl AB  \\
5028.126  & $-$1.02  & K:RDLB & 28819&954 &  &bl B &  & bl A \\
5068.766  & $-$1.042 & K:FW   & 23711&456 &  &bl AB  & & bl B \\
5098.698  & $-$2.03  & K:FW   & 17550&181 & &bl AB, red comp ?  &  & bl AB \\
5107.447  & $-$3.087 & K:FW   &  7985&785 &  &bl A & & bl B \\
5107.641  & $-$2.418 & K:FW   & 12560&934 & $-$4.57      &bl A &   $-$4.27 & bl B \\
5110.413  & $-$3.760 & K:FW   &     0&000 &   &bl A, core &   & bl B \\
5126.192  & $-$1.06  & K:FW   & 34328&752 &   & bl B, unk, redsh ? &  &bl A, unk ? \\
5137.381  & $-$0.43  & K:FW   & 33695&397 & & bl AB, redsh  & & bl B \\
5142.494  & $-$0.739 & K:K17  & 34692&148 &  & bl AB, redsh  & & bl AB, core \\
5142.540  & $-$0.295 & K:K17  & 34328&752 &  $-$4.57 & bl AB, redsh & $-$4.27 & bl AB, core    \\
5196.059  & $-$0.477 & K:K17  & 34328&752 &  & bl AB, redsh ?& & bl AB \\
5216.274  & $-$2.150 & K:FW   & 12968&554 &  & bl AB, red comp ?& & bl AB \\
5217.389  & $-$1.07  & K:DRLP & 25899&989 &  & bl AB, redsh ?&  $-$4.57& single \\
5232.940  & $-$0.057 & K:FW   & 23711&456 &  $-$4.67 & almost single& & bl A \\
5364.870  & $+$0.228 & K:FW   & 35856&402 &  $-$4.67 & single& $-$4.57& single \\
5367.465  & $+$0.443 & K:FW   & 35611&625 &  $-$4.97 & almost single, core&  $-$4.87& almost single, core \\
5369.961  & $+$0.536 & K:FW   & 35257&324 &  $-$4.77 & almost single&   & bl A \\
5371.489  & $-$1.645 & K:FW   &  7728&060 &          & bl AB &  & bl AB \\
5373.708  & $-$0.71  & K:RDLB & 36079&372 &          & bl AB &    & bl B \\
5383.368  & $+$0.645 & K:FW   & 34782&421 &          & bl B &  $-$4.57& almost single \\
5389.478  & $-$0.430 & K:K17  & 35611&625 &  $-$4.67 & almost single &  $-$4.47& almost single \\
5400.501  & $-$0.151 & K:K17  & 35257&324 &           & bl AB  &     & bl AB \\
5405.774  & $-$1.844 & K:FW   &  7985&785 &           & bl AB  &     & bl A \\
5410.909  & $+$0.398 & K:FW   & 36079&372 &           & bl AB, redsh, red comp ?& $-$4.47& almost single \\
5415.198  & $+$0.642 & K:FW   & 35279&308 &           & bl AB & $-$4.67& almost single \\
5424.067  & $+$0.780 & K:K17  & 34843&957 &           & bl A, core &    & bl AB, core \\
5429.696  & $-$1.879 & K:FW   &  7728&060 &           & bl A &         & bl AB \\
5445.042  & $+$0.209 & K:K17  & 35379&208 &  $-$4.82 &single, core  &        & bl AB \\
5446.916  & $-$1.914 & K:FW   &  7985&785 &           &bl AB  &      & bl AB \\
5466.395  & $-$0.630 & FMW    & 35257&324 &           &bl A  &        & bl A \\
5487.745  & $-$0.316 & K:K17  & 34843&957 &  $-$4.67 & single, core  &  & bl AB \\
5497.516  & $-$2.849 & K:FW   &  8154&714 &          & bl A  &      & bl B, core \\
5506.779  & $-$2.797 & K:FW   &  7985&785 &          & bl A  &      & bl B \\
5522.446  & $-$1.52  & K:FW   & 33946&933 &          & bl B  &      & bl A \\
5560.212  & $-$1.16  & K:FW   & 35767&564 &  $-$4.47 & single, core&  & bl A \\
5572.842  & $-$0.28  & K:DRLP & 27394&691 &          & bl A, core  &  & bl AB \\
5914.111  & $-$0.362 & K:K17  & 37162&764 &          & bl A, core  &   & bl B \\
5914.201  & $-$0.111 & K:K17  & 37162&764 &  $-$4.47      & bl A, core  & $-$4.67 & bl B \\
6191.557  & $-$1.416 & K:FW   & 19621&006 &   & bl A &             & bl B \\
6393.600  & $-$1.58  & K:FW   & 19621&006 & $-$4.37 & single, redsh 1\,km\,s$^{-1}$ ?, comp ? &  $-$4.27& single \\
\hline
\end{tabular}
\end{table*}

\setcounter{table}{0}

\begin{table*}
\caption[ ]{cont.}
\centering
\begin{tabular}{lclr @{.} lcl|cl}
\hline\noalign{\smallskip}
\multicolumn{1}{c}{$\lambda^{a}$ [\AA{}]} &
\multicolumn{1}{c}{$\log gf$}&
\multicolumn{1}{c}{Ref.$^{b}$}&
\multicolumn{2}{c}{$\chi_{\rm low}$}&
\multicolumn{1}{c}{$\log \frac{N_{\rm Z}}{N_{\rm tot(A})}$}&
\multicolumn{1}{l}{Notes$^{c}$}&
\multicolumn{1}{c}{$\log \frac{N_{\rm Z}}{N_{\rm tot(B})}$}&
\multicolumn{1}{l}{Notes$^{c}$}\\
\hline\noalign{\smallskip}
\multicolumn{1}{c}{\ion{Fe}{ii}} & & & \multicolumn{2}{c}{} & $-$4.42 &$-$4.37 (Sun) &$-4.44\pm0.06$ & $-$4.57 (Sun)  \\
\hline\noalign{\smallskip}
4178.854  & $-$2.535 & RU2    & 20830&582 &        & bl B, unk &  & bl A \\
4413.591  & $-$3.985 & SUN    & 21581&638 &  $-$4.42 & single &     & very bl\\
4416.819  & $-$2.601 & RU2    & 22409&852 &  & bl B &  & bl A\\
4515.533  & $-$2.540 & RU2    & 22409&852 & & bl AB, redsh&  & bl AB\\
4620.513  & $-$3.188 & K:FW   & 22810&357 &  & bl B &  & bl A, redsh ?\\
4635.317  & $-$1.474 & K:FW   & 48039&090 &  & bl B &  & bl A \\
4923.921  & $-$1.206 & K:FW   & 23317&635 &  & bl AB &  & bl AB \\
5100.655  & $-$4.222 & K:FW   & 22637&195 &   & bl B        &  $-$4.32 & bl \ion{Fe}{ii}\\
5120.344  & $-$4.275 & K:K13  & 22810&357 &   & bl AB&  &bl AB\\
5132.661  & $-$4.008 & K:FW   & 22637&205 &  & bl B  &  &bl A\\
5197.568  & $-$2.229 & K:FW   & 26055&412 &  & bl A, red wing, unk ? &  &bl A\\
5362.869  & $-$2.616 & RU2    & 25805&327 &  & bl AB&      & bl AB\\ 
5425.249  & $-$3.352 & K:FW   & 25805&327 &  & bl B & $-$4.47 & single \\
5534.838  & $-$2.86  & K:FW   & 26170&18  & $-$4.42 & single & $-$4.47 & single \\
6084.102  & $-$3.854 & K:FW   & 25805&327 &      & bl B &   & bl unk ? \\
6147.734  & $-$2.731 & K:FW   & 31364&455 &        & bl A, \ion{Fe}{i} &  & bl AB\\
6149.246  & $-$2.732 & K:FW   & 31368&453 &        & bl B & $-$4.47 & single\\
6383.730  & $-$2.275 & K:FW   & 60445&279 &        & bl A, \ion{Fe}{i} &  & bl AB\\
6456.380  & $-$2.086 & K:FW   & 31483&198 &        & bl B & $-$4.47 & single\\
\hline\noalign{\smallskip}
\multicolumn{1}{c}{Fe(tot)} & & & \multicolumn{2}{c}{} & $-4.62\pm0.16$ &$-$4.57 (Sun) & $-4.49\pm0.15$ &$-$4.57 (Sun) \\
\hline
\hline\noalign{\smallskip}
\multicolumn{1}{c}{\ion{Co}{i}} & & & \multicolumn{2}{c}{} & ---  &$-$7.11 (Sun) &$-$6.91 & $-$7.11 (Sun)  \\
\hline\noalign{\smallskip}
3995.302h & $-$0.220 & NIST5  &  7442&410 &  & bl AB & $-$6.91 & single \\ 
4118.767h & $-$1.093 & K:K08  &  8460&783 &  & bl AB &         & bl A \\
4121.311h & $-$0.320 & NIST5  &  7442&410 &  & bl B &           & bl A   \\ 
\hline\noalign{\smallskip}
\multicolumn{1}{c}{\ion{Ni}{i}} & & & \multicolumn{2}{c}{} & $-5.84\pm0.08$  &$-$5.84 (Sun) &$-5.70\pm0.17$ & $-$5.84 (Sun)  \\
\hline\noalign{\smallskip}
4295.882h & $-$0.480 & NIST5  & 30979&749 &  & bl AB &  & bl AB\\ 
4648.652  & $-$0.150 & NIST5  & 27580&391 &  & bl B   &  & bl A  \\
4701.530  & $-$0.390 & NIST5  & 32973&376 &  & bl AB &  & bl AB \\
4756.515  & $-$0.270 & NIST5  & 28068&065 & $-$5.84 & almost single&  & bl A\\ 
4806.987  & $-$0.640 & NIST5  & 29668&893 &  $-$5.84 & single &  & bl A\\ 
4829.023h & $-$0.330 & NIST5  & 28569&203 & $-$5.84 & almost single&  $-$5.84 &almost single \\ 
4831.176  & $-$0.410 & NIST5  & 29084&450 &  $-$5.94 & single &  & bl A \\ 
4904.412h & $-$0.170 & NIST5  & 28569&203 &  & bl B & $-$5.84 & single  \\ 
5035.362h & $+$0.290 & NIST5  & 29320&762 &  $-$5.94 & single & $-$5.84 & single \\
5035.967h & $-$0.234 & K88    & 29480&989 &   & bl A &  & bl B  \\
5081.110  & $+$0.300 & NIST5  & 31031&020 &  & bl B &  & bl A \\ 
5082.344h & $-$0.540 & NIST5  & 29500&674 &  & bl B &  & bl A \\ 
5085.480h & $-$1.480 & NIST5  & 29500&674 &  &  bl B &$-$5.84 & single \\ 
5099.920  & $-$0.100 & NIST5  & 29668&893 &  & bl B & $-$5.74 & single \\ 
5102.960h & $-$2.620 & NIST5  & 13521&347 & $-$5.94 &  single &   & bl A \\ 
5115.392  & $-$0.110 & NIST5  & 30922&734 &  & bl B &  & bl A \\ 
5146.482h & $+$0.060 & K:K08  & 29888&477 &  & bl A &  & bl B \\ 
5155.764  & $-$0.090 & NIST5  & 31441&635 &  & bl B &  $-$5.44 & single\\ 
5578.726  & $-$2.640 & NIST5  & 13521&347 & $-$5.74 & single & $-$5.44 & single, spike, cont\\
6643.543h & $-$2.300 & NIST5  & 13521&347 & $-$5.74 & single & $-$5.59 & single\\
6772.315  & $-$0.990 & NIST5  & 29500&674 & $-$5.74 & single &  & bl A\\
\hline\noalign{\smallskip}
\multicolumn{1}{c}{\ion{Cu}{i}} & & & \multicolumn{2}{c}{} & $-$7.86& $-$7.86 (Sun) &$-$7.86  & $-$7.86 (Sun) \\
\hline\noalign{\smallskip}
5105.548  & $-$1.500 & NIST5  & 11202&565 & $-$7.86: & bl B & $-$7.86 & single\\
5153.238  & $+$0.116 & K:K12  & 30535&302 &  & bl B &  & bl A\\
5218.202  & $+$0.264 & NIST5  & 30783&686 &  & bl A &  & bl A\\
\hline
\noalign{\smallskip}
\end{tabular}
\end{table*}

\setcounter{table}{0}

\begin{table*}
\caption[ ]{cont.}
\centering
\begin{tabular}{lclr @{.} lcl|cl}
\hline\noalign{\smallskip}
\multicolumn{1}{c}{$\lambda^{a}$ [\AA{}]} &
\multicolumn{1}{c}{$\log gf$}&
\multicolumn{1}{c}{Ref.$^{b}$}&
\multicolumn{2}{c}{$\chi_{\rm low}$}&
\multicolumn{1}{c}{$\log \frac{N_{\rm Z}}{N_{\rm tot(A})}$}&
\multicolumn{1}{l}{Notes$^{c}$}&
\multicolumn{1}{c}{$\log \frac{N_{\rm Z}}{N_{\rm tot(B})}$}&
\multicolumn{1}{l}{Notes$^{c}$}\\
\hline\noalign{\smallskip}
\multicolumn{1}{c}{\ion{Zn}{i}} & & & \multicolumn{2}{c}{} & $-$7.38:& $-$7.48 (Sun) &$-$7.58  & $-$7.48 (Sun) \\
\hline\noalign{\smallskip}
4680.136  & $-$0.810 & K:K12  & 32311&319 &         & bl AB &  & bl AB\\
4722.157  & $-$0.338 & K:K12  & 32501&330 &         & bl B &  & bl A\\
4810.532  & $-$0.125 & K:K12  & 32890&327 & $-$8.08? &single, bad cont ? & $-$8.08 ? & single, bad cont ?\\
6362.346  & $+$0.160 & NIST5  & 46771&199 & $-$7.38: & bl B &  $-$7.58 & single\\
\hline\noalign{\smallskip}
\multicolumn{1}{c}{\ion{Sr}{i}} & & & \multicolumn{2}{c}{} & ---   &$-$9.21 (Sun)  &$-$9.0  & $-$9.21 (Sun)  \\
\hline\noalign{\smallskip}
4607.333  & $+$0.283 & NIST5  &     0&000 &  & bl B &  $-$9.00 & single\\
\hline\noalign{\smallskip}
\multicolumn{1}{c}{\ion{Sr}{ii}} & & & \multicolumn{2}{c}{} & $-$9.10 &$-$9.21 (Sun) &$-$8.90  & $-$9.21 (Sun) \\
\hline\noalign{\smallskip}
4077.709  & $+$0.148 & NIST5  &     0&000 &           &bl AB  &       & bl AB \\
4215.519  & $-$0.173 & NIST5  &     0&000 &   $-$9.1  &bl A &  $-$8.9& bl AB  \\
4305.443  & $-$0.11  & NIST5  & 24516&65  &        & bl AB &       & bl B  \\
\hline\noalign{\smallskip}
\multicolumn{1}{c}{Sr(tot)} & & & \multicolumn{2}{c}{} & $-$9.1  &$-$9.21 (Sun) & $-8.95\pm0.05$ &$-$9.21 (Sun) \\
\hline
\hline\noalign{\smallskip}
\multicolumn{1}{c}{\ion{Y}{ii}} & & & \multicolumn{2}{c}{} & $-$9.40 &$-$9.83 (Sun) &$-$9.50  & $-$9.83 (Sun)  \\
\hline\noalign{\smallskip}
4177.530h & $-$0.163 & NIST5  &  3296&180 &  &bl AB &  & bl B\\
4309.622  & $-$0.747 & NIST5  &  1449&752 &  &bl A &  & bl B\\
4374.933h & $+$0.155 & NIST5  &  3296&184 &  &bl AB &  & bl AB\\
4398.010  & $-$0.999 & NIST5  &  1045&083 & &bl A &   &bl A \\
4854.861h & $-$0.38  & NIST5  &  8003&126 & & bl B &   & bl B \\
4883.682h & $+$0.07  & NIST5  &  8743&316 & $-$9.40 & single &  & bl A, cont ?\\ 
4900.119h & $-$0.09  & NIST5  &  8328&041 &          &cont ?, bl A &  &cont ?, bl B\\ 
5087.419h & $-$0.17  & NIST5  &  8743&316 & $-$9.40 & single & $-$9.50 & bl A\\
5119.112h & $-$1.36  & NIST5  &  8003&121 &  & bl A &  & bl AB\\
5205.723h & $-$0.35  & NIST5  &  8328&039 &  & bl AB &  &  bl B \\ 
5509.895h & $-$1.015 & NIST5  &  8003&126 &  & bl A   &  & bl B \\ 
\hline\noalign{\smallskip}
\multicolumn{1}{c}{\ion{Zr}{ii}} & & & \multicolumn{2}{c}{} & $-$9.04   &$-$9.45 (Sun) &$-$9.45  & $-$9.45 (Sun)  \\
\hline\noalign{\smallskip}
4149.198  & $-$0.040 & Ljun   &  6467&610 &   & bl A &    & bl AB   \\ 
4156.232  & $-$0.780 & Ljun   &  5724&380 &   & bl unk &   & bl unk   \\ 
4208.980  & $-$0.510 & Ljun   &  5752&920 &  $-$9.04 & single &   $-$9.45 & single   \\ 
4258.041  & $-$1.200 & Ljun   &  4505&500 &   & bl A &    & bl AB   \\ 
4442.992  & $-$0.420 & Ljun   & 11984&460 &   & bl AB &  & bl AB   \\ 
4496.962  & $-$0.890 & Ljun   &  5752&92  &  & bl A &  & bl AB   \\ 
5112.270  & $-$0.850 & Ljun   & 13428&500 &  $-$9.04 & single &  & bl A   \\ 
\hline\noalign{\smallskip}
\multicolumn{1}{c}{\ion{Ba}{i}} & & & \multicolumn{2}{c}{} & &$-$9.79 (Sun)&  & $-$9.79 (Sun)  \\
\hline\noalign{\smallskip}
5535.481  & $+$0.215 & NIST5  &     0&000 &  & bl A   &  & bl AB \\
\hline\noalign{\smallskip}
\multicolumn{1}{c}{\ion{Ba}{ii}} & & & \multicolumn{2}{c}{} & $-$9.51   &$-$9.79 (Sun) &$-$9.41  & $-$9.79 (Sun)  \\
\hline\noalign{\smallskip}
4130.649  & $+$0.524 & NIST5  & 21952&36  &  & bl A &  & bl A \\ 
4554.033h & $+$0.140 & NIST5  &     0&000 &    $-$9.51 &almost single &  $-$9.41 & almost single \\
4934.077h & $-$0.157 & NIST5  &     0&000 &  & bl AB &  & bl A\\
5853.675  & $-$0.908 & NIST5  &  4873&852 &  & bl B &  & bl A \\
6141.710  & $-$0.032 & NIST5  &  5674&807 &  & bl A &   & bl B  \\
6496.898h & $-$0.407 & NIST5  &  4873&852 &  & bl telluric&  & bl telluric  \\
\hline
\end{tabular}
\end{table*}

\setcounter{table}{0}

\begin{table*}
\caption[ ]{cont.}
\centering
\begin{tabular}{lclr @{.} lcl|cl}
\hline\noalign{\smallskip}
\multicolumn{1}{c}{$\lambda^{a}$ [\AA{}]} &
\multicolumn{1}{c}{$\log gf$}&
\multicolumn{1}{c}{Ref.$^{b}$}&
\multicolumn{2}{c}{$\chi_{\rm low}$}&
\multicolumn{1}{c}{$\log \frac{N_{\rm Z}}{N_{\rm tot(A})}$}&
\multicolumn{1}{l}{Notes$^{c}$}&
\multicolumn{1}{c}{$\log \frac{N_{\rm Z}}{N_{\rm tot(B})}$}&
\multicolumn{1}{l}{Notes$^{c}$}\\
\hline\noalign{\smallskip}
\multicolumn{1}{c}{\ion{La}{ii}} & & & \multicolumn{2}{c}{} & $-10.22\pm0.15$   &$-$10.93 (Sun) &$-$10.57  & $-$10.93 (Sun) \\
\hline\noalign{\smallskip}
3988.515h & $+$0.210 & NIST5  &  3250&35  &  & bl A &    & bl A   \\ 
3995.745  & $-$0.064 & NIST5  &  1394&46  &   & cont &   & bl AB   \\ 
4042.901  & $+$0.270 & L:CB'  &  7473&32  & $-$10.37 & single &  & bl A \\ 
4086.709h & $-$0.070 & NIST5  &     0&000 &  & bl B  &  $-$10.57 & single \\ 
4123.218h & $+$0.130 & NIST5  &  2591&60  & & bl AB  &   & bl B \\ 
4333.75   & $-$0.059 & NIST5  &  1394&46  & $-$10.07 & single  &   & bl A \\ 
\hline\noalign{\smallskip}
\multicolumn{1}{c}{\ion{Ce}{ii}} & & & \multicolumn{2}{c}{} & $-$10.46   & $-$10.46 (Sun) & ---  & $-$10.46 (Sun) \\
\hline\noalign{\smallskip}
4186.596  & 0.813    & K:MC   &  6967&547 &  & bl A  &   & bl B \\ 
4562.358  & 0.210    & NIST5  &  3854&012 &  & bl B &  & bl A\\
4628.161  & 0.140    & NIST5  &  4165&55  &  $-$10.46 & single &  & bl A\\
\hline\noalign{\smallskip}
\multicolumn{1}{c}{\ion{Nd}{ii}} & & & \multicolumn{2}{c}{} & $-$10.62 ? & $-$10.62 (Sun)&$-$10.62 ? & $-$10.62 (Sun) \\
\hline\noalign{\smallskip}
4061.080  & $+$0.550 & NIST5  &  3801&93  &$-$10.62   & bl A & $-$10.62 & bl B \\ 
4109.071  & $-$0.163 & NIST5  &   513&330 & $-$9.98 & bl AB &$-$9.98  & bl AB\\
4303.571  & $+$0.084 & NIST5  &     0&000 & $-$10.62 &bl B  &  $-$10.62 & bl A \\
4706.543  & $-$0.710 & NIST5  &     0&000 &  $-$9.92  & bl B & $-$9.92 & single \\ 
5076.580  & $-$0.250 & K:MC'  &  5985&59  & $>$ $-$10.62  & bl A &   $-$10.62 & bl B \\ 
\hline
\end{tabular}
\end{table*}

\bsp	
\label{lastpage}
\end{document}